\documentclass[fleqn,usenatbib]{mnras}
\usepackage{newtxtext,newtxmath}
\usepackage{booktabs}

\usepackage[T1]{fontenc}

\DeclareRobustCommand{\VAN}[3]{#2}
\let\VANthebibliography\thebibliography
\def\thebibliography{\DeclareRobustCommand{\VAN}[3]{##3}\VANthebibliography}

\usepackage{graphicx}	
\usepackage{float} 
\usepackage{amsmath}	
\usepackage{subcaption}
\usepackage{tabularx}

\title[The Fate of Exomoons Around Hot Jupiters]{The Fate of Exomoons Around Hot Jupiters}

\author[Pu et al.]{
Yangjun Pu,$^{1,2}$\thanks{E-mail: alnitakpu@gmail.com}
Chenyang Li,$^{1,2}$
Bohang Zhu$^{1,2}$
and Huigen Liu$^{1,2,3}$\thanks{E-mail: huigen@nju.edu.cn}
\\
$^{1}$School of Astronomy and Space Science, Nanjing University, Nanjing, 210046, China\\
$^{2}$Key Laboratory of Modern Astronomy and Astrophysics, Ministry of Education, Nanjing, 210046, China\\
$^{3}$MEAYIN Science Center, Nanjing, 210046, China
}

\date{Accepted XXX. Received YYY; in original form ZZZ}

\pubyear{\the\year{}}

\begin{document}
\label{firstpage}
\pagerange{\pageref{firstpage}--\pageref{lastpage}}
\maketitle

\begin{abstract}
Despite extensive searches, no exomoon has been confirmed. Cold gas giants are expected to host moons, while hot Jupiters (HJs), which migrate inward from distant formation sites, provide dynamically hostile environments for moons. In this paper, we systematically investigate the ultimate fate of exomoons around gas giants undergoing two primary migration mechanisms: disk migration and coplanar high-eccentricity migration (CHEM). Through $N$-body simulations using \textsc{rebound}, we find that during disk migration, retrograde moons exhibit a significantly higher survival fraction ($\sim 25\%$) compared to prograde ones ($\sim 8\%$). Conversely, for systems that successfully evolve into hot Jupiters via CHEM, no moons can survive the extreme orbital excitation; they are inevitably engulfed or gravitationally stripped before the inward migration. Interestingly, massive exomoons ($\sim 10\,M_{\oplus}$) can exert a strong dynamical feedback on the host planet, acting as a dynamical shield that completely halts the high-eccentricity migration process to form HJs in $\sim 30\%$ of cases, preserving the satellite system. Furthermore, during chaotic multi-body interactions, $\sim 10\%$ of gas giants are dynamically ejected as free-floating planets (FFPs), with roughly $40\%$ of these ejected planets successfully retaining their moons. Combining these dynamical instabilities with severe physical barriers such as atmospheric stripping and magnetic torques, we conclude that mature hot Jupiters are generally devoid of observable primordial moons. Consequently, future observational missions aiming to detect exomoons should focus on cold Jupiters and FFPs, which offer a more stable environment for moon retention.
\end{abstract}




\section{Introduction}
Moons are ubiquitous in our Solar System, especially around giant planets like Jupiter and Saturn. The discovery and characterization of exomoons around exoplanets could offer profound insights into their distinct formation mechanisms such as giant impacts or accretion within circumplanetary disks \citep{2004ARA&A..42..441C,2012Icar..219..737M}.
Furthermore, large exomoons orbiting gas giants in the habitable zone may possess liquid water on their surfaces, making them potentially habitable environments for life \citep[e.g.,][]{1997Natur.385..234W,2010ApJ...712L.125K}.
Although exomoons are theorized to be abundant, their observational confirmation remains challenging, and definitive detections have yet to be made \citep{2024arXiv241006248S}.

Currently, transit photometry and transit timing variations (TTVs) serve as the most viable techniques for detecting these elusive companions \citep{2009MNRAS.392..181K, 2013ApJ...777..134K}.
Utilizing data from the Kepler Space Telescope, \citet{2013ApJ...777..134K} conducted a comprehensive search for exomoons.
However, to date, only two promising candidates have emerged: Kepler-1625 b-i and Kepler-1708 b-i \citep{2022NatAs...6..367K,2018SciA....4.1784T}, with estimated radii of roughly 5 and 2.6 $R_{\oplus}$, respectively.
Due to observational biases favoring the transit detection of exceptionally large moons, these isolated discoveries do not necessarily confirm the widespread existence of massive moons in the universe.
Moreover, the physical reality of these signals remains a subject of intense debate; recent studies suggest the observed variations could instead be attributed to planetary rings or dynamical perturbations from unseen planets \citep{2024NatAs...8..193H}.

Beyond cold Jupiters, the potential for exomoons orbiting hot Jupiters (HJs) has also been extensively investigated.
Typically defined by \citet{2018ARA&A..56..175D} as planets with masses exceeding $0.25 M_{\mathrm{J}}$ and orbital periods under 10 days, HJs are generally considered unlikely to retain moons.
Tidally, the equilibrium point of the star–planet–moon system for HJs often either does not exist or lies beyond the stable Hill sphere \citep{2016MNRAS.462.2527A}. 
Magnetically, moons orbiting HJs usually reside within the corotation radius, causing them to spiral inward and cross the planetary Roche limit on Myr timescales \citep{2024ApJ...965...88W}.
In addition, intense stellar irradiation in such close-in environments can drive severe photo-evaporation, which has been shown to induce planetary mass loss and destabilize circumplanetary moon systems around close-in planets \citep{2016ApJ...833....7Y}.
Consequently, mature hot Jupiters-moon system lack long-term dynamical stability.

Current theoretical consensus proposes two primary pathways for HJ formation: disk migration and high-eccentricity migration \citep{2018ARA&A..56..175D}. 
In the disk migration model, a massive planet opens a gap in the protoplanetary disk and exchanges angular momentum with the surrounding gas, experiencing negative Lindblad torques.
This interaction drives the planet inward from an initial orbit of $1$-$5\,\mathrm{AU}$ to within $0.1\,\mathrm{AU}$ before the disk dissipates \citep{1980ApJ...241..425G,1986ApJ...309..846L,2014prpl.conf..667B}.
Conversely, in the high-eccentricity migration scenario, giant planets undergo severe eccentricity excitation through mechanisms such as the Lidov-Kozai mechanism, secular chaos, planet-planet scattering, or coplanar interactions \citep{1962AJ.....67R.579K,1962P&SS....9..719L,1989Natur.338..237L,2004ApJ...604..388I,2015ApJ...805...75P}.
The highly eccentric planet subsequently loses orbital energy via tidal dissipation at pericentre, ultimately circularizing its orbit to become a hot Jupiter \citep{1998ApJ...499..853E}.

Considering these distinct formation mechanisms, \citet{2010ApJ...719L.145N} demonstrated that the timescale for planetary tides to affect moon orbits is much longer than the timescale of disk migration. 
Therefore, a planet is likely to become hot Jupiter well before its outer moon migrates inside the final Hill Sphere due to tides. 
For moons in close orbits, \citet{2002ApJ...575.1087B} argued that low-mass, difficult-to-detect moons are more dynamically stable than their higher-mass counterparts.
Although subsequent studies employing alternative tidal models suggest that even Earth-mass moons in close orbits might maintain stable over Gyr timescales \citep{2009ApJ...704.1341C}, these investigations primarily consider already-formed systems and neglect the dynamical disturbance during the inward migration process itself.

In contrast, the high-eccentricity migration mechanism, particularly the initial phase of extreme eccentricity excitation, is highly detrimental to moon stability.
Planet-planet scattering mechanism is generally a destructive process, rendering the resulting HJs almost incapable of retaining moons \citep{2013ApJ...769L..14G}. 
Specifically, moons with semi-major axes exceeding 0.1 Hill radii ($R_{\mathrm{H}}$) are systematically removed from the system \citep{2018ApJ...852...85H}. 
Under the Lidov-Kozai mechanism, \citet{2020MNRAS.499.4195T} found that moons cannot survive the excitation phase.
Interestingly, massive Neptune-sized moons possess a $\sim 10\%$ chance of suppressing the planet's eccentricity excitation entirely \citep{2020MNRAS.499.4195T}.

Although extensive research suggests that mature hot Jupiters are unlikely to retain exomoons, the precise evolutionary pathways and dynamical fates of these primordial moons during the migration process remain incompletely understood. 
Previous literature has primarily focused on the long-term tidal and magnetic decay of already-formed HJs over Myr timescales.
Furthermore, studies examining moon survivability during violent migration processes frequently overlook retrograde moons, despite evidence showing that retrograde orbits are significantly more robust against external perturbations than prograde ones.

Importantly, the specific mode of eccentricity excitation critically dictates the survival rate of these moons.
The coplanar mechanism, for instance, can excite planetary eccentricities rapidly without the need for violent close planetary encounters.
This relatively gentle and swift excitation drastically reduces the severity of disruptive perturbations compared to the Lidov-Kozai mechanism and E2 mechanism \citep{2017ApJ...848...20W}. 
Therefore, this study aims to systematically investigate the stability and final fates of exomoons with varying masses and inclinations during both disk migration and coplanar high-eccentricity migration. 
Utilizing the \textsc{rebound} N-body code with the IAS15 integrator \citep{2012A&A...537A.128R,2015MNRAS.446.1424R}, we place special emphasis on massive retrograde moons, assessing their orbital resilience against external perturbations and quantifying how their strong dynamical feedback dictates the ultimate fate of the host planet.

The remainder of this paper is organized as follows: Section 2 describes the numerical methods and adopted models of disk migration and coplanar high-eccentric migration. 
Section 3 presents the simulation results regarding exomoon stability in disk migration and CHEM scenario. 
Section 4 provides an analytical formulation of the moon-shielding effect, compares our results with previous literature, and discusses additional physical mechanisms that prevent hot Jupiters from retaining moons.
Finally, Section 5 summarizes the major conclusions.

\section{Methods}

\subsection{Disk Migration Model}\label{sec:2.1}
Planets with masses greater than that of Saturn can carve a gap in the protoplanetary disk. 
The giant planet becomes trapped in this annular gap; gravitational torques exerted by both the inner and outer regions of the disk confine the planet, causing it to migrate inward on the same timescale as the viscous evolution of the gas.
The process is known as Type II disk migration \citep{1986ApJ...309..846L,2014prpl.conf..667B}. 

We adopt the Type II migration rate limited by the disk's viscous transport, given by \citet{2004ApJ...604..388I,2005A&A...434..343A}:
\begin{equation}\label{eq1}
    \left(\frac{\mathrm{d} a}{\mathrm{d} t}\right)_{\mathrm{II}} = -\frac{3 \nu}{2 a} \min \left(1, 2 \Sigma_{\mathrm{p}} \frac{a^{2}}{M_{\mathrm{p}}}\right)
\end{equation}
where $M_{\mathrm{p}}$ is the planetary mass and $\Sigma_P$ is the column density of the disk nearby the orbit. 
The disk gas column density follows a power-law profile $\Sigma_{\mathrm{p}}=\Sigma_{\mathrm{p}, 0}* a^{-3/2}$, in which $\Sigma_{\mathrm{p}, 0}$ is $10^3 g/cm^2$ at 1$\mathrm{\ AU}$, representing a classic value \citep{1997ApJ...486..372B}.
The kinematic viscosity is $\nu=\alpha c_{\mathrm{s}}^2/n$, where ${c_{s}}=\sqrt{k_{\mathrm{B}} T / (\mu m_{\mathrm{p}})}$ is the isothermal sound speed, $k_{\mathrm{B}}$ is the Boltzmann constant, $m_{\mathrm{p}}$ is the proton mass, and $n$ is the Keplerian orbital frequency at radius $a$.
The kinematic viscous efficiency parameter of the disk $\alpha$ is sampled from $10^{-3}$ to $10^{-2}$, in accordance with observations and simulations  \citep{2007MNRAS.376.1740K,2019NewA...70....7M}, and we adopt a mean molecular weight of $\mu=1.85$.

It should be noted that the primary objective of this work is not to investigate the general probability or universality of hot Jupiter formation via disk migration, but rather to construct a robust and continuous dynamical pathway that transports gas giants into the inner system to examine the subsequent fate of their exomoons.
Thus, we adopt an integrated framework for the disk thermodynamics, disk torques and the stellar tides following \citet{2019A&A...628A..42H}. I.e. the local mid-plane temperature of the disk is calculated by:
\begin{equation}\label{eq2}
    T=\left[\frac{3}{2}\left(\frac{\tau}{2}+\frac{1}{\sqrt{3}}+\frac{1}{3 \tau}\right) \frac{k_{\mathrm{B}}}{\sigma_{\mathrm{SB}}} \frac{\alpha}{\mu} \Sigma_{\mathrm{p}} n\right]^{1 / 3}
\end{equation}
In Equation~(\ref{eq2}), the optical depth is defined as $\tau=\kappa \Sigma_{\mathrm{p}} / 2$, where the Rosseland mean opacity $\kappa$ is sampled between $10^{-7}$ and $10^{-6} \mathrm{\ cm^{2} \ g^{-1}}$. 

In addition, Type II migration ceases in our model when the planet reaches the inner 2:1 mean motion resonance with the radial position of the magnetic cavity of the star ($r_{\mathrm{mc}}$).
We adopt values for $r_{\mathrm{mc}}$ ranging from 0.055 to 0.08 $\mathrm{AU}$, which are broadly consistent with the estimations by \citet{2002ApJ...574L..87K}. 
Consequently, these parameters cause gas giants to halt their migration at 0.035 to 0.05$\mathrm{\ AU}$, matching the typical semi-major axes of hot Jupiters.
To reduce computational costs during extensive N-body integrations, the orbital decay rate $\dot{a}$ was artificially enhanced by an acceleration factor $\eta = 30$, while ensuring that the migration timescale remains significantly longer than the orbital periods of both the planet and its moon, thereby preserving the adiabatic nature of the dynamical evolution.

Meanwhile, stellar tides can result in strong orbital dissipation for planets close to their host stars.
In general, planets with orbital semi-major axes larger than the stellar corotation radius $r_{\mathrm{co}}$
are repelled, whereas planets interior to $r_{\mathrm{co}}$ will gradually lose their orbital energies.
This mechanism creates a tidal migration barrier and can explain the concentration of hot Jupiters at $\sim 0.04\mathrm{\ AU}$ \citep{2019A&A...628A..42H}.

Consistent with our integrated methodology, we adopt the star-planet equilibrium tidal dissipation formulism derived by \citet{1981A&A....99..126H,1998ApJ...499..853E} and 
\begin{equation}\label{eq3}
\begin{aligned}
\left(\frac{\mathrm{d} a}{\mathrm{d} t}\right)_{\mathrm{tidal}} &= -a \sum_{i=\star, \mathrm{p}} \frac{1}{T_{i}}\left[\frac{f_{1}(e)}{\beta^{15}}-\frac{f_{2}(e)}{\beta^{12}} \frac{\Omega_{i}}{n}\right] \\
\left(\frac{\mathrm{d} e}{\mathrm{d} t}\right)_{\mathrm{tidal}} &= -e \sum_{i=\star, \mathrm{p}} \frac{9}{2 T_{i}}\left[\frac{f_{3}(e)}{\beta^{13}}-\frac{11 f_{4}(e)}{18 \beta^{10}} \frac{\Omega_{i}}{n}\right],
\end{aligned}
\end{equation}

We direct the reader to Equation (7) through (17) in \citet{2019A&A...628A..42H} for the precise expression of $\beta$ and $T_{i},f_{1},f_{2},f_{3}$ and $f_{4}$.
In our model, the stellar spin period is uniformly distributed between 2 and 8 days, which aligns with observations of young low-mass stars in NGC 2362 and Taurus \citep{2008MNRAS.384..675I,2020AJ....159..273R}.
Similarly, the planetary spin period is drawn uniformly from 8 to 24\,h, representative of typical rotation states of gas giants.

Exomoons orbiting gas giants could migrate inward or outward during planetary migration due to planet-moon tidal interaction. 
However, the timescale of Type-II migration is typically much shorter than that of the tidal evolution of moons' orbit \citep{2010ApJ...719L.145N} .
Therefore, the semi-major axis of the moon remains nearly stationary relative to the planet during disk migration, eventually leading to exomoons' collision with the planet or its ejection from the system when the Hill radius shrinks below the exomoon's orbital radius.
We incorporated the formalism proposed by \citet{2017MNRAS.471.3019A} in our test runs, and the results demonstrate that the tidal effects on moons are negligible within the rapid disk migration scenario. 
As such, we do not account for planet-moon tidal forces in our primary simulations.

Changes in the planetary radius could also alter the fate of moons during planets' migration.
An abrupt radial contraction or change in rotation could push moons outward, mitigating the likelihood of them being engulfed by the planet \citep{2017MNRAS.471.3019A}. 
However, planets undergoing disk migration usually do not experience this process. 
Since \citet{2021ApJ...909L..16T} claimed that gas giants cool and contract slowly, meanwhile the process of disk migration is rapid, the amplitude of planetary radial contraction during this phase is negligible. Therefore, we assume a constant planetary radius in our simulations.

\subsection{Coplanar High-eccentricity Migration Model}\label{sec2.2}

In the high-eccentricity migration scenario, the gas giant's eccentricity is excited by various mechanisms, after which it undergoes orbital decay through tidal dissipation. 
Here we focus mainly on the formation of hot Jupiters via coplanar high-eccentricity migration \citep{2015ApJ...805...75P}.
In this channel, the eccentricity of an inner gas giant is excited by a more massive, highly eccentric outer planet through secular gravitational interactions, followed by a gradual decrease in its semi-major axis and eccentricity due to stellar tides acting on the planet.
Due to the nearly coplanar configuration of the system, the resulting hot Jupiters typically possess low orbital inclinations \citep{2015ApJ...805...75P}. 
Such high eccentricities for the outer perturber can naturally arise from stellar flybys in young star clusters \citep{2013A&A...549A..82P,2019MNRAS.490...21H}. 
These close encounters can directly perturb the gas giant or excite distant companions, which subsequently transfer their angular momentum deficit (AMD) to the inner planet through secular interactions.

We adopt the first typical initial configuration proposed by \citet{2015ApJ...805...75P}, because of its higher efficiency in exciting eccentricity compared to their second configuration. The initial requirements can be expressed as:
\begin{equation}\label{eq4}
    e_{p} \leq 0.1, \quad e_{2} \geq 0.67, \quad \frac{M_{p}}{M_{2}}\left(\frac{a_{p}}{a_{2}}\right)^{1/2} \leq 0.3
\end{equation}
where $e$, $M$, $a$ are eccentricity, mass, semi-major axis separately. For the subscripts, p refers to the inner planet and 2 refers to the outer.

At extreme eccentricities, tidal energy dissipation is highly localized near pericentre, raising a substantial tidal bulge on the giant planet. 
This impulsive deformation strongly excites internal oscillation modes within the planet, a mechanism known as the dynamical tide \citep{1995ApJ...450..722M,1995ApJ...450..732M}. 
For highly eccentric orbits, the energy dissipated by these dynamical tides is orders of magnitude stronger than that of standard equilibrium tides. 
Therefore, we exclusively adopt the dynamical tidal model for our simulations.
To implement this in a secular framework, we time-average the intense, episodic energy dissipation occurring at pericentre over the entire orbital period. Accordingly, the evolution rates of the planet's semi-major axis and eccentricity are derived by \citet{1981A&A....99..126H,2017AJ....154..272H,2018ApJ...854...44M,2022ApJ...931...11G}:

\begin{equation}\label{eq5}
\begin{aligned}
\left(\frac{\mathrm{d} a}{\mathrm{d} t}\right)_{\mathrm{dyn}} &= -\frac{2 f_{\mathrm{dyn}} R_{\mathrm{p}}^{8} M_{\star} n^{2}}{G M_{\mathrm{p}}^{2} \mathcal{T} a^{4}} A(e) e^{2} \frac{f(e)}{\left(1-e^{2}\right)^{15/2}} \\
\left(\frac{\mathrm{d} e}{\mathrm{d} t}\right)_{\mathrm{dyn}} &= -\frac{f_{\mathrm{dyn}} R_{\mathrm{p}}^{8} M_{\star} n^{2}}{G M_{\mathrm{p}}^{2} \mathcal{T} a^{4}} A(e) e \frac{f(e)}{\left(1-e^{2}\right)^{13/2}}
\end{aligned}
\end{equation}
where the auxiliary enhancement functions $A(e)$ and $f(e)$ for highly eccentric orbits are defined as \citet{2017AJ....154..272H}:
\begin{equation}\label{eq6}
\begin{aligned}
A(e) &\equiv \frac{\left(1-e^{2}\right)^{15/2}}{(1-e)^{9} e^{2} f(e)} \\
f(e) &:= \frac{1+\frac{45}{14} e^{2}+8 e^{4}+\frac{685}{224} e^{6}+\frac{255}{448} e^{8}+\frac{25}{1792} e^{10}}{1+3 e^{2}+\frac{3}{8} e^{4}}
\end{aligned}
\end{equation}
Here, $\mathcal{T}$ is the orbital period of the planet (where $\mathcal{T} = 2\pi/n$). The dimensionless scaling factor $f_{\mathrm{dyn}}$ characterizes the efficiency of the dynamical tide. In our work, we adopt $f_{\mathrm{dyn}} = 0.5$ following \citep{1986ApJ...306..552M}.

We deliberately adopt the dynamical tides instead of the equilibrium tides for several reasons.
Firstly, dynamical tides are overwhelmingly dominant for extremely eccentric orbits.
For instance, for a gas giant at $1 \mathrm{\ AU}$ with an eccentricity of 0.98, dynamical tides are approximately 1500 times more powerful than equilibrium tides according to Eq. \ref{eq3} and \ref{eq5}. I.e. planets driven by dynamical tides can experience an orbital decay of $\sim10^{-4}\mathrm{\ AU}$ each year.
In contrast, equilibrium tides would only reduce $a_{\mathrm{p}}$ by $\sim 10^{-7} \mathrm{\ AU}$ per orbit, a dissipation rate even weaker than the secular perturbations exerted by the outer planet.
Therefore, if the inner planet is to migrate inwards and rapidly decouple from the outer planet, its eccentricity generally must be excited above 0.99.
However, this extreme excitation frequently results in the planet falling inside the stellar Roche radius or being ejected from the system, even in non-coplanar configurations.

Secondly, although the two tidal models yield similar orbital evolution during the low-eccentricity stages, our focus remains on the final phase of excitation and migration process. 
The initial period of the simulation is relatively inconsequential to the final fate of planets and moons, meaning that replacing equilibrium tides with dynamical tides throughout the simulation introduces negligible error.

Lastly, resolving the extremely slow migration driven solely by equilibrium tides would dramatically increase the integration time, making our large-scale simulations computationally prohibitive. 
The specific orbital parameters and masses explored in our N-body simulations will be detailed in Section 3.

\section{Results}

\subsection{Disk Migration}\label{sec:3.1}

In the disk migration scenario, each simulation is initiated with a Jupiter-mass planet ($M_{\mathrm{p}} = 1\ M_{\mathrm{J}}$) located at $1 \ \mathrm{AU}$ with an initial planetary radius of $R_{\mathrm{p}} = 2\ R_{\mathrm{J}}$.
We terminate the migration once the planet reaches the inner disk boundary $0.63r_{\mathrm{mc}}$, and subsequently completes 10,000 orbits around the star.

Starting from $1\ \mathrm{AU}$, the inward migration of the gas giant to a hot Jupiter (with an orbital period of $\approx 3\ \mathrm{days}$) takes approximately $10^5\ \mathrm{yr}$ ($0.1\ \mathrm{Myr}$). 
Upon reaching the magnetic truncation radius, the planet's orbital migration stalls, as the positive repulsive tidal torque counterbalances the vanishing disk torque in the inner cavity.
The specific location of the magnetic cavity ultimately dictates the final semi-major axis of the hot Jupiter.
While the overall migration rate is governed by the disk viscosity parameter $\alpha$, the shorter migration timescale within $1\,\mathrm{AU}$ naturally reflects the shorter local viscous timescale of the inner disk.
For comparison, benchmark simulations initialized at $5\,\mathrm{AU}$ yield a migration timescale of $\sim 1.5\,\mathrm{Myr}$, consistent with canonical Type-II migration models and the typical lifetimes of protoplanetary disks.

\subsubsection{Fate of Single Moon through Disk Migration}\label{sec:3.1.1}

In our N-body simulations, the initial mass of the moons is sampled as log-uniform distribution between $10^{-12}$ and $10^{-4}$ solar mass, assuming a constant bulk density of $4g/cm^{3}$.
For the initial orbital configurations, the semi-major axis $a_{\mathrm{m}}$ of prograde moons is uniformly distributed between $0.03$ and $0.055$ times reduced Hill radius ($R_{\mathrm{h}}$, see Appendix \ref{appendix A}), whereas for retrograde moons, it ranges from $0.015R_{\mathrm{h}}$ to $0.055R_{\mathrm{h}}$. It is worthwhile to mention that initial location of moons is distinct for prograde and retrograde scenarios respectively, because of the different stability around planets, see in Appendix \ref{appendix A}.
The initial eccentricity for all moons is uniformly set to  $e = 0.001$.
The inner boundary of the initial semi-major axis is chosen to reside just above the planetary surface, while the outer limit is strategically set as the migrating planet's reduced Hill radius. Since
the gas giant planets are highly inflated, their physical radii can extend beyond the corresponding Roche limits. 

Depending on the position of magnetic cavity, the hot Jupiter stalls its migration at $a_{\mathrm{p}} = 0.038 - 0.05 \ \mathrm{AU}$. This final orbital architecture creates a narrow survival belt bounded by the shrinking $R_{\mathrm{H}}$ and the inflated planetary radius. The radial widths of this theoretical stable zone are severely restricted, spanning to several times of planet radii depending on the prograde or retrograde orbits.

\begin{figure*}
    \centering
    \hspace*{-20mm}
    \includegraphics[width=0.8\textwidth]{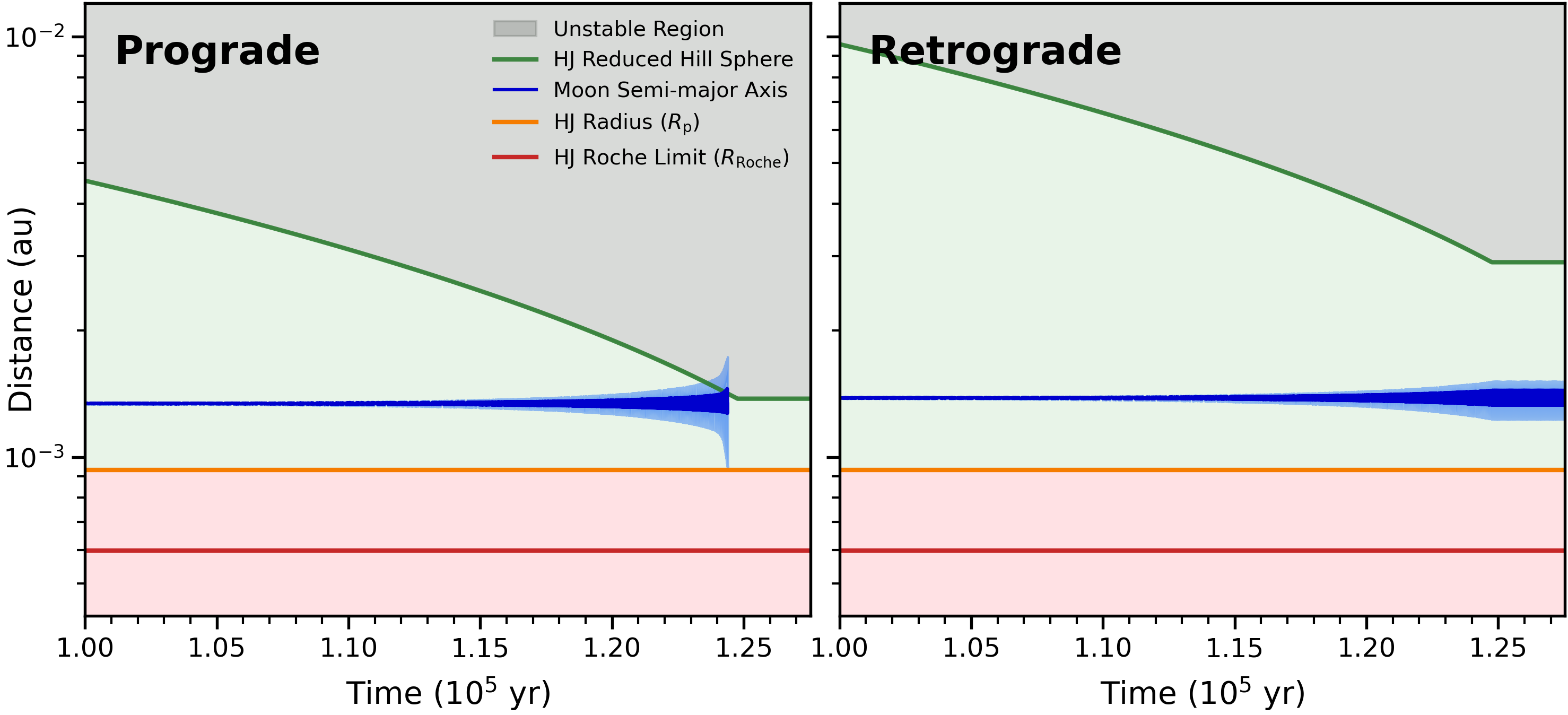}
    \caption{Orbital evolution of a prograde (left) and a retrograde (right) exomoon placed at a distance of $2 \times 10^{5}$ kilometers ($1.43R_{p}$) from the center of the migrating hot Jupiter during the final stage of disk migration process. Parameters include $\alpha = 0.003$, $\kappa = 10^{-6}$, and $R_{\mathrm{mc}} = 0.07 \ \mathrm{AU}$. The solid blue line tracks the moon's semi-major axis while the shaded blue region indicates the radial distance variation from pericentre to apocentre. The green line represents the stability boundary (reduced Hill radius). The orange and red lines denote the planetary radius ($R_{\mathrm{p}}$) and the Roche limit ($R_{\mathrm{Roche}}$), respectively. The light green area highlights the theoretically stable zone for the moon, bounded by the dynamically unstable outer region (grey) and the tidal disruption zone (pink).}
    \label{fig1}
\end{figure*}

Fig.~\ref{fig1} illustrates the orbital evolution of both a prograde and a retrograde moon, initially placed at a distance of $2 \times 10^{5}$ kilometers ($1.43R_{p}$) from the center of the migrating hot Jupiter.
It is evident that while the moons generally maintain their initial semi-major axes throughout the migration phase, their eccentricities are susceptible to excitation driven by stellar gravitational perturbations.
Particularly for prograde moons, even if their semi-major axes remain nominally within the reduced Hill radius, their orbital architectures are frequently disrupted as the gas giant reaches its final close-in orbit.
The excitation of eccentricity inevitably drives the prograde moons into extreme orbital oscillations, either pushing them beyond the planet's gravitational control or plunging them into the deep atmospheric envelope.
In contrast, retrograde moons residing within the stable belt exhibit higher survival possibilities, stemming from the wider stability zone and their intrinsically more robust dynamical configuration against external perturbations. 

\begin{figure*}
    \centering
    \includegraphics[width=0.9\textwidth]{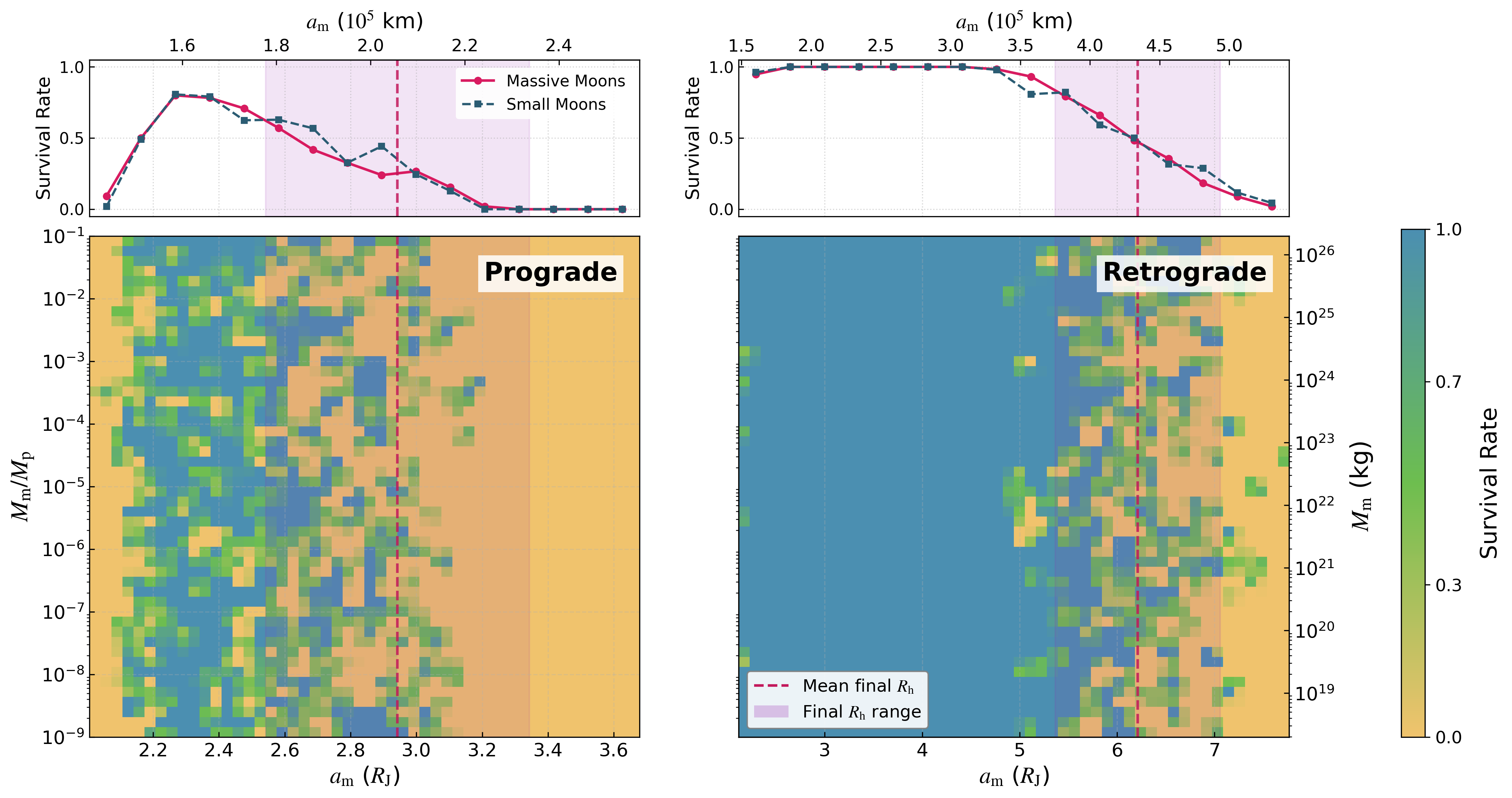}
    \caption{Survival rate maps and one-dimensional radial distributions for a single prograde (left) and retrograde (right) exomoon following Type-II disk migration. \textbf{Bottom panels:} Survival fraction heatmaps as a function of the moon's initial semi-major axis ($a_{\mathrm{m}}$) and mass ($M_{\mathrm{m}}$). The color map indicates the empirical survival probability, ranging from 0 (golden; orbital instability, ejection, or collision) to 1 (blue; long-term circumplanetary retention). \textbf{Top panels:} Binned survival rates along the semi-major axis (16 bins), comparing massive moons ($M_{\mathrm{m}} \ge 10^{-5}\,M_{\mathrm{p}}$, solid magenta line with circles) and small moons ($M_{\mathrm{m}} < 10^{-5}\,M_{\mathrm{p}}$, dashed dark-blue line with squares). In all panels, the red vertical dashed line denotes the mean final reduced Hill radius ($R_{\mathrm{h}}$) of the hot Jupiter, while the faint purple shaded range illustrates the possible range of the final $R_{\mathrm{h}}$ arising from uncertainties in the planetary parking location ($a_{\mathrm{p}} = 0.038 - 0.050$\,AU).}
    \label{fig2}
\end{figure*}

We simulated 4800 groups of gas giant disk migration with a prograde moon or a retrograde moon, respectively, and plotted the relationship between moon survival rate and its mass or initial semi-major axis in Fig.~\ref{fig2}. 
It is clear that retrograde moons orbiting hot Jupiters formed by disk migration have a higher chance of survival.

For prograde moons, if the stable region is defined as the domain where a moon possess a survival probability exceeding $50\%$, its radial width is remarkably narrow, spanning merely $6\times10^{4} \mathrm{km}$.
Specifically, this prograde survival zone is bounded between an inner limit of approximately $1.5 \times 10^5 \ \mathrm{km} \ (1.07R_{p})$ and an outer limit of $2.1 \times 10^5 \ \mathrm{km} \ (1.5R_{p})$ from the center of the hot Jupiter. 
Dynamically, the stable zone for retrograde moons is vastly more expansive. 
Its outer boundary extends up to $4.3 \times 10^5 \ \mathrm{km}$, which precisely reflects the larger theoretical stability coefficient (i.e. the reduced hill radius $R_{\mathrm{h}}=0.95 \ R_{\mathrm{H}}$, see Appendix \ref{appendix A}) intrinsic to retrograde orbits.
Remarkably, the location of this simulated outer boundary aligns perfectly with the theoretical critical stability limit for a prograde moon ($0.45 \ R_{\mathrm{H}}$) or a retrograde moon ($0.95 \ R_{\mathrm{H}}$)around the final hot Jupiter. 

Furthermore, as quantitatively revealed by the top panels of Fig.~\ref{fig2}, the survival rate profiles along the semi-major axis are virtually indistinguishable between the massive ($M_{\mathrm{m}} \ge 10^{-5}\,M_{\mathrm{p}}$) and small ($M_{\mathrm{m}} < 10^{-5}\,M_{\mathrm{p}}$) sub-samples. 
This directly demonstrates that the retention probability in single-moon systems is largely insensitive to the moon's mass. 
This feature is dynamically consistent, as the Hill stability boundary primarily depends on the mass ratio of the primary binary (star-planet), rendering circumplanetary orbits effectively independent of the third body's mass in the test-particle regime.
This uniform survival behavior across a wide mass spectrum holds true for both prograde and retrograde orbits, ranging from Hyperion-mass to Sub-Neptune-mass objects.

Our simulations also indicate that an extended planetary migration distance (e.g., migrating from $5 \ \mathrm{AU}$ instead of $1 \ \mathrm{AU}$) does not enhance the moon's survival prospects.
Because the orbital evolution driven by planetary migration drastically outpaces the local tidal orbital decay, the ultimate fate of the moon is predominantly dictated by its initial orbital parameters.
Assuming a logarithmically uniform distribution for the moon's initial semi-major axes within the circumplanetary disk (representing a neutral baseline prior in the absence of formation constraints), the overall survival rate for prograde moons is merely $7.9\%$ for a gas giant migrating from $1 \ \mathrm{AU}$ and it further drops to $5.7\%$ for one migrating from $5 \ \mathrm{AU}$.
In sharp contrast, retrograde moons exhibit significantly higher survival rates of $24.5\%$ and $17.9\%$, respectively.
This reduction in survival rate for planets migrating from further out occurs because cold Jupiters at $5 \ \mathrm{AU}$ initially possess larger gravitational spheres of influence, allowing moons to form at greater distances. 
As the planet migrates inward and its Hill sphere drastically shrinks, these distant moons are inevitably stripped away.

Additionally, as summarized in Table~\ref{tab1}, a small fraction ($4.3\%$) of retrograde moons in single-satellite systems are neither engulfed nor retained.
Instead, as the planet's Hill Sphere progressively contracts, these companions can be gently detached from circumplanetary orbits into independent heliocentric trajectories.
This transitions them into a class of liberated bodies termed ploonets \citep{2025A&A...694L...8S}.
Because retrograde orbits possess wider stability thresholds and opposite angular momentum relative to the planetary motion, detached moons can smoothly decouple across the outer Lagrange points rather than directly plunging into the planetary envelope.
Once detached, these bodies become independent heliocentric planets. A similar pathway
was identified by \citet{2026ApJ..1000..202G} for moons around migrating
circumbinary planets, where $38\%$ of the simulated moons became
long-period circumbinary ploonets. Their subsequent evolution is,
however, sensitive to the planetary architecture and disk properties.
The ploonets in the single-planet circumbinary systems of
\citet{2026ApJ..1000..202G} generally remained on wide orbits over the
$1\,\mathrm{Myr}$ disk lifetime, those formed in our compact
multi-planet systems are expected to cross orbits with the migrating
giant, triggering secondary encounters that lead to planetary
collisions, ejections or stellar engulfment.

\subsubsection{Fate of Multiple Moons through Disk Migration}\label{sec:3.1.2}

Given that giant planets in the Solar System typically host multiple moons, such as Jupiter's four Galilean moons, it is necessary to investigate the survivability of multi-moon systems during planetary disk migration.
To this end, we simulated 2400 disk migration cases featuring multiple prograde or retrograde moons. 
In each simulation, the system comprises three initially coplanar moons of equal mass.
The initial semi-major axis distribution of the innermost moon is identical to that used in our single-moon simulation, while the adjacent moons are randomly separated by $80$ to $200$ mutual reduced Hill radii ($R_{\mathrm{h,m}}$) for prograde systems, and $40$ to $120 \ R_{\mathrm{h,m}}$ for retrograde systems.
This spacing criterion not only ensures the dynamical stability of the moons at the beginning of the simulation, but also mimics a realistic formation scenario where the circumplanetary disk provides sufficient material to sustain their growth.
All other physical parameters remain consistent with the single-moon disk migration models.

\begin{figure*}
    \centering
    \includegraphics[width=0.9\textwidth]{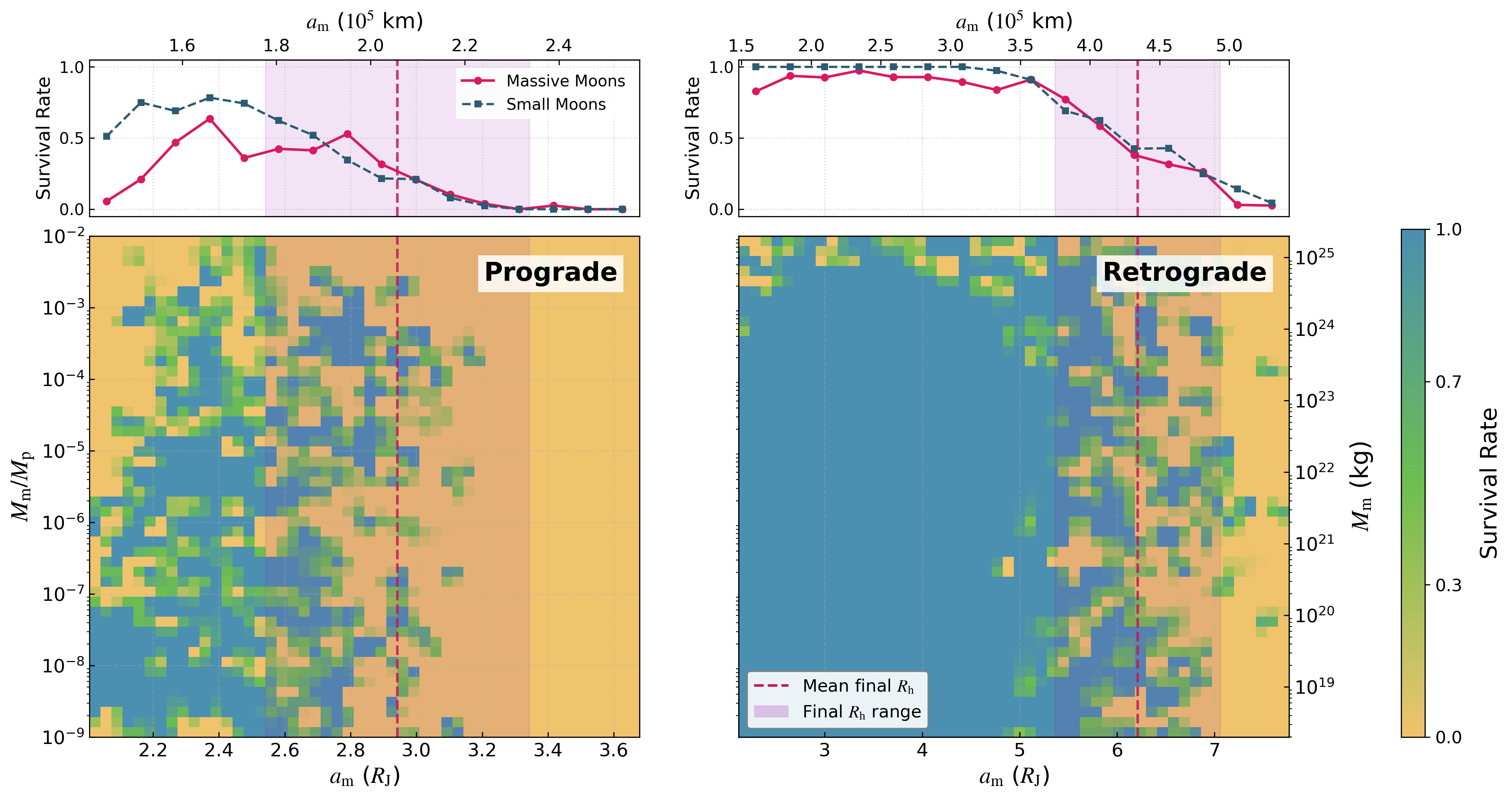}
    \caption{Same as Figure.~\ref{fig2}, but for multi-moon systems comprising three initially coplanar moons of equal mass. \textbf{Bottom panels:} System-level survival rate heatmaps, where a system is deemed to have survived if at least one of the three moons remains bound on a stable orbit after disk migration. \textbf{Top panels:} One-dimensional radial distributions of the system-level survival rate (16 bins), separated by the moon mass threshold $M_{\mathrm{split}} = 10^{-5}\,M_{\mathrm{p}}$ into massive moons (solid magenta line) and small moons (dashed dark-blue line). Note the pronounced suppression in the retention of massive prograde systems caused by mutual moon-moon scattering.}
    \label{fig3}
\end{figure*}

Fig.~\ref{fig3} shows the relationship between the survival rate of these multi-moon systems and their initial mass and semi-major axis, with the integrated statistical outcomes summarized in Table~\ref{tab1}.
We define a system as survival system with stable moon if at least one of the three placed moons is retained in a stable orbit, without distinguishing the exact number of surviving moons.
Comparing the multiple-moon outcomes (Fig.~\ref{fig3}) with the single-moon results (Fig.~\ref{fig2}), we observe that the survival rate of retrograde moons decreasing with $a_{moon}$ exhibits no significant deviations.
However, the stability of massive prograde moons is compromised, even though the theoretical width of their stability zone remains unchanged. 

As quantitatively revealed in the top panels of Fig.~\ref{fig3}, this divergence is primarily driven by intense moon-moon scattering.
For massive moon systems, their relatively large gravitational reach causes the outer companions to become dynamically unstable before the planet finalizes its migration into a hot Jupiter.
For prograde architectures, because their available stable zone is intrinsically narrow, this disruption is catastrophic: within the nominal stable belt ($a_{\mathrm{moon}} \approx 1.6 - 2 \times 10^5$\,km), the retention fraction of massive prograde systems drops precipitously from $\sim 75\%$ down to $\sim 40\%$ compared to their low-mass counterparts, and even plunges to nearly zero at the innermost boundary.
Conversely, for retrograde massive moons, this dynamic disruption is not fatal, as their substantially wider stability zone allows them to absorb moderate orbital excitations without crossing instability boundaries.
Throughout the wide interior stability plateau ($a_{\mathrm{moon}} \lesssim 3.5 \times 10^5$\,km), massive retrograde moon systems maintain an exceptionally high survival probability exceeding $85\%$, suffering only marginal suppression.
In the outer transition regime ($a_{\mathrm{moon}} > 3.5 \times 10^5$\,km), both mass cohorts trace nearly identical decline profiles, demonstrating that external Hill-sphere contraction governs the outer truncation boundary.

\begin{table*}
\centering
\caption{Dynamical Fates and Retention Fractions of Exomoons during Type-II Disk Migration}
\label{tab1}
\small
\begin{tabular*}{\textwidth}{@{\extracolsep{\fill}}lcccc}
\hline\hline
 & \multicolumn{2}{c}{\textbf{Single-Moon Systems}} & \multicolumn{2}{c}{\textbf{Multi-Moon Systems (3 Moons)}} \\
\cmidrule(lr){2-3} \cmidrule(lr){4-5}
Dynamical Outcome & Prograde Moons & Retrograde Moons & Prograde Moons & Retrograde Moons \\
\hline
\textbf{Moon Fate} & & & & \\
Stable Moon & 31.6\% & 71.9\% & 13.2\% & 37.9\% \\
Moon Collided on Star & 0.0\% & 0.0\% & 0.0\% & 0.0\% \\
Moon Collided on Planet & 68.4\% & 23.8\% & 64.0\% & 32.4\% \\
Moon Escaped & 0.0\% & 0.1\% & 0.0\% & 0.1\% \\
Moon Became Planet (Ploonet) & 0.0\% & 4.3\% & 22.8\% & 29.5\% \\
\hline
\textbf{System-Level Retention} & & & & \\
At Least One Moon Survived$^a$ & 31.6\% & 71.9\% & 29.5\% & 68.3\% \\
\hline
\textbf{Extrapolated Full-Disk Survival Fraction}$^b$ & 7.9\% & 24.5\% & Severely degraded & Marginally degraded \\
\hline\hline
\end{tabular*}
\begin{flushleft}
\footnotesize
$^a$ Survival probability defined by retaining at least one bound circumplanetary moon after the host planet completes disk migration and circularization. \\
$^b$ Overall fraction integrated assuming a primordial log-uniform distribution of moons across the circumplanetary disk for migration originating at 1\,AU, as detailed in Section~3.1.1.
\end{flushleft}
\end{table*}

Furthermore, as cataloged in Table~\ref{tab1}, mutual dynamical scattering efficiently ejects outer companions from the circumplanetary domain, decoupling roughly 22.8\% of prograde and 29.5\% of retrograde moons into independent heliocentric ploonets.
However, these detached ploonets predominantly represent a transient dynamical state. 
In the presence of the gaseous disk, such low-mass objects undergo rapid Type~I migration alongside strong eccentricity damping, outside the gap due to gravity of giant planets, a timescale far shorter than the host planet's viscous Type~II migration.
Consequently, we speculate that these ploonets will inevitably cross orbits with the migrating giant, triggering secondary close encounters that culminate in planetary collisions or stellar engulfment.

\subsection{Coplanar High-eccentricity Migration}\label{sec:3.2}

\subsubsection{Dynamical Evolution of the Planets}\label{sec:3.2.1}

For the high eccentricity migration scenario, we initialize the inner gas giant (the hot Jupiter progenitor) on a nearly circular orbit ($e_{\mathrm{p}} = 0.1$) at $1\mathrm{\ AU}$, and the outer perturbing giant on a highly eccentric orbit ($e_{\mathrm{2}} \in \{0.7, 0.8\}$) with a semi-major axis $a_{\mathrm{2}}$ chosen from $\{5.5, 6.0, 6.5, 7.0\} \mathrm{\ AU}$ .
The mass of the inner planet is fixed at $1 M_{\mathrm{J}}$, while the outer perturber's mass is randomly set to either $2 M_{\mathrm{J}}$ or $5 M_{\mathrm{J}}$, ensuring that all selected parameter combinations robustly satisfy equation~(\ref{eq4}). 
To represent a roughly coplanar architecture, the mutual inclinations of both planets are drawn randomly from a uniform distribution between $-9^{\circ}$ to $9^{\circ}$ relative to the reference plane. 
The remaining orbital elements including true anomaly ($f$), longitude of the ascending node ($\Omega$), and argument of periapsis ($\omega$)---are assigned completely randomly.
The radius of the inner planet is $1.5R_{\mathrm{J}}$. We set a single moon around the inner planet. The moon masses are sampled log-uniformly from $3\times10^{18}\,\mathrm{kg}\ (10^{-9}\,M_{\mathrm{p}})$ to $3\times10^{26}\,\mathrm{kg}\ (10^{-1}\,M_{\mathrm{p}})$; while the initial semi-major axes ($a_{\mathrm{m}}$) are randomly distributed between $0.03R_{\mathrm{h}}$ and $0.11R_{\mathrm{h}}$ for prograde moons or $0.015 R_{\mathrm{h}}$ and $0.055 R_{\mathrm{h}}$ for retrograde moons. The initial eccentricity is fixed at $0.01$, and we simulate 1200 cases for both prograde and retrograde moons. 

The specific initial orbital configuration implies that upon strong eccentricity excitation, the orbits of the two giant planets may inevitably cross.
We deliberately adopt this compact architecture for two primary purposes.
Firstly, the extremely short orbital period of a close-in exomoon makes long-term integrations computationally prohibitive; placing the perturber relatively close accelerates the secular excitation and substantially reduces the required simulation time.
Secondly, the intersecting orbits sometimes trigger strong planet-planet scattering events, providing a valuable opportunity to concurrently investigate the survival rates of exomoons around free-floating planets (FFPs).

The numerical simulations are terminated when any of the following four conditions are met:
(1) the inner planet successfully undergoes migration and moderate circularization to form a hot Jupiter (defined as $a_{\mathrm{p}} < 0.1 \mathrm{\ AU}$ and $e_{\mathrm{p}} < 0.2$);
(2) the inner planet collides with the host star;
(3) the inner planet is ejected from the system via scattering, becoming an FFP (defined as $a_{\mathrm{p}} < 0$, which means a hyperbolic orbit, and the distance from the star $> 100\mathrm{\ AU}$);
(4) the inner planet fails to achieve high-eccentricity excitation.

To optimize the allocation of computational resources, the "failure to excite" condition is enforced via a maximum wall-clock time limit of 8 hours per simulation case.
Additionally, we implement a progressive early-stopping algorithm based on the physical evolution timescale, where a simulation is prematurely aborted if the maximum planetary eccentricity $e_{\max}$ fails to reach 0.3 within 10,000 years, 0.5 within 30,000 years, 0.7 within 50,000 years, 0.9 within 70,000 years, or 0.97 within 100,000 years. 

\begin{figure}
    \centering
    \includegraphics[width=\linewidth]{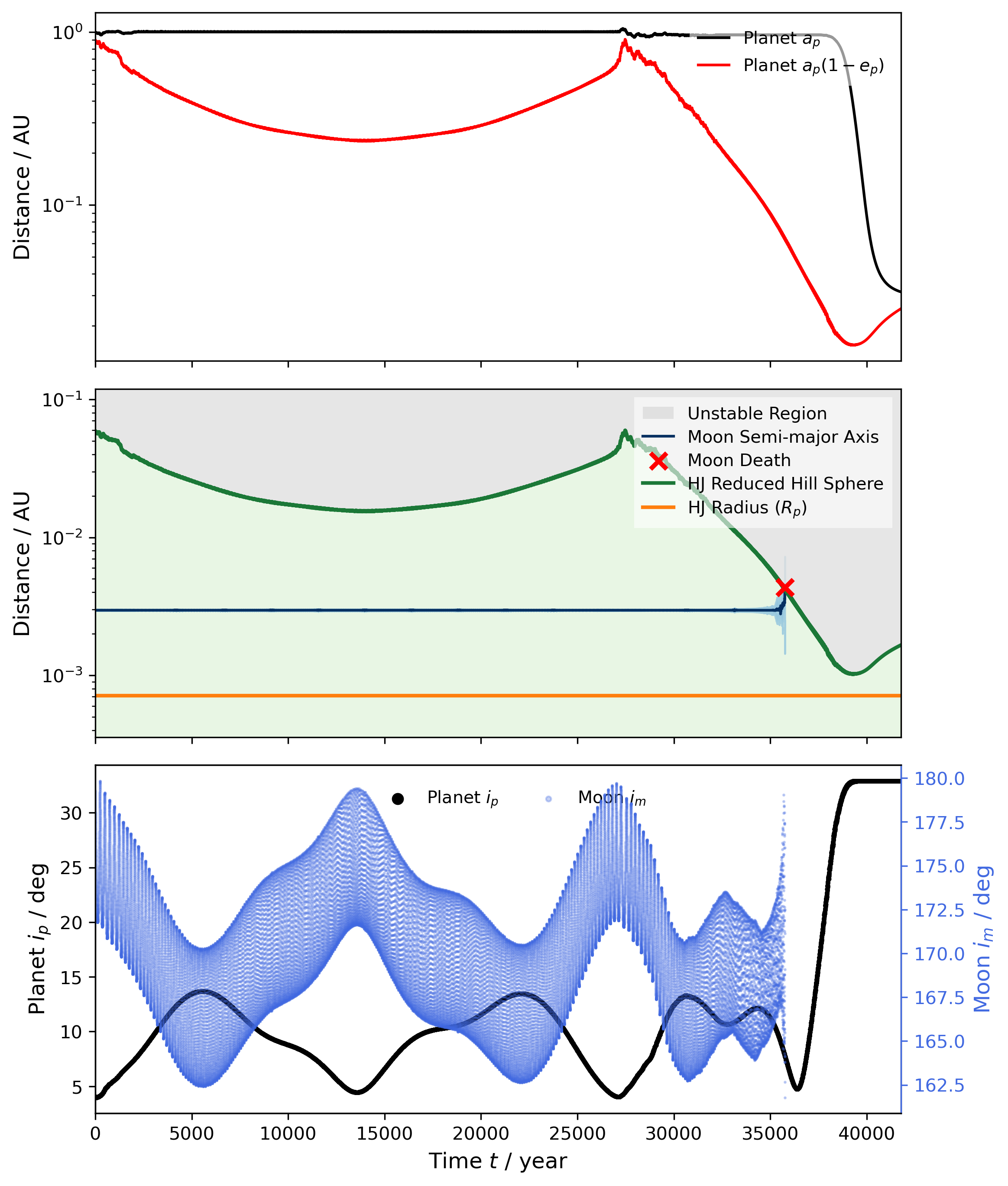}
    \caption{\textbf{Orbital evolution and stability map of a retrograde exomoon during planetary high-eccentricity migration.} 
            \textit{Top panel:} Evolution of the host planet's semi-major axis ($a_p$, black curve) and pericentre distance ($a_p(1-e_p)$, red curve). The drastic plunge in pericentre distance illustrates the intense orbital decay during the high-eccentricity migration phase. 
            \textit{Middle panel:} The dynamical survival space of the exomoon. The dark blue curve tracks the moon's semi-major axis ($a_m$), which is strictly constrained by the theoretical stability limit (green curve). The shaded grey area represents the unstable region, while the orange horizontal line denotes the physical radius of the host planet ($R_p = 1.5 R_{\mathrm{J}}$). The red cross marks the exact epoch when the moon is gravitationally stripped from the collapsing Hill sphere, escaping planetary bound orbit and transitioning into an independent heliocentric ploonet. 
            \textit{Bottom panel:} Inclination evolution of the planet ($i_p$, black dots, left axis) and the retrograde moon ($i_m$, blue dots, right axis). The double y-axis configuration effectively highlights the intense nodal precession and dynamical coupling between the planet and the moon prior to the moon's detachment.}
    \label{fig4}
\end{figure}

To clearly visualize this migration pathway, Fig.~\ref{fig4} depicts the eccentricity excitation and subsequent tidal circularization in an exemplary system.
In this case, the inner planet is excited by an outer perturber ($m_2 = 5\,M_{\mathrm{J}}$, $a_2 = 6\,\mathrm{AU}$, $e_2 = 0.7$).
During the initial $35,000\,\mathrm{yr}$, the inner planet undergoes pure angular momentum exchange with the perturber, driving its eccentricity to an extreme peak of nearly $0.985$ while its orbital energy (and thus semi-major axis $a_p$, top panel) remains conserved.
Once the pericentre distance plunges sufficiently close to the central star, intense dynamical tides rapidly dissipate orbital energy, causing $a_p$ to drop precipitously from $1\,\mathrm{AU}$ down to a compact orbit within merely $\sim 10,000\,\mathrm{yr}$.
With our adopted tidal framework, the planet achieves a maximum migration rate of up to $2 \times 10^{-4} \mathrm{\ AU}$ per year.
This rapid pacing demonstrates that coplanar secular migration is a highly efficient channel for producing a vast population of hot Jupiters.
Furthermore, such a short evolutionary timescale ensures that the $N$-body integrations remain computationally affordable.
Additionally, the bottom panel tracks the inclination evolution, demonstrating that the newly formed hot Jupiter maintains a moderate inclination of $30^\circ$.

As shown in Table \ref{tab2}, approximately half of the migrating gas giants successfully achieving this compact circularized configuration within merely $100$ kyr.
Meanwhile, a substantial fraction of the planets (roughly $40\%$) remain unexcited as cold Jupiters, indicating that their initial perturbative conditions are insufficient to trigger extreme eccentricity growth.
Beyond these bound outcomes, violent multi-body scattering events result in the complete ejection of approximately $8\%$ to $11\%$ of the gas giants, transforming them into FFPs.
Additionally, a marginal fraction of the planets (roughly $1.0\%$) directly collide with the central star, a catastrophic fate driven either by intense chaotic scattering or excessively rapid eccentricity excitation that overwhelms the tidal circularization process.

\subsubsection{Unstable Moons Around Hot Jupiters}\label{sec:3.2.2}

\begin{table*}
    \centering
    \begin{tabular}{lcccc}
        \hline
        & \multicolumn{2}{c}{Prograde Moons} & \multicolumn{2}{c}{Retrograde Moons} \\
        \textbf{Fate of System} & Large Moons & Small Moons & Large Moons & Small Moons \\
        \hline
        \textbf{Hot Jupiters (Total)} & \textbf{52.58\%} & \textbf{52.94\%} & \textbf{48.21\%} & \textbf{52.90\%} \\
        \quad -- Stable Moon & 0.00\% & 0.00\% & 0.00\% & 0.00\% \\
        \quad -- Moon Collided on Star & 6.36\% & 7.82\% & 4.26\% & 6.63\% \\
        \quad -- Moon Collided on Planet & 37.80\% & 37.30\% & 34.58\% & 34.83\% \\
        \quad -- Moon Escaped & 7.90\% & 7.33\% & 8.01\% & 9.62\% \\
        \quad -- Moon Became Planet (Ploonet) & 0.52\% & 0.49\% & 1.36\% & 1.82\% \\
        \hline
        \textbf{Escaped Jupiters / FFPs (Total)} & \textbf{10.14\%} & \textbf{7.97\%} & \textbf{10.73\%} & \textbf{10.63\%} \\
        \quad -- with Stable Moon & 3.61\% & 3.91\% & 5.28\% & 4.48\% \\
        \quad -- Moon Collided on Star & 0.69\% & 0.16\% & 0.17\% & 0.50\% \\
        \quad -- Moon Collided on Planet & 3.26\% & 3.09\% & 3.24\% & 4.48\% \\
        \quad -- Moon Escaped & 1.03\% & 0.81\% & 1.70\% & 1.00\% \\
        \quad -- Moon Became Planet (Ploonet) & 1.55\% & 0.00\% & 0.34\% & 0.17\% \\
        \hline
        \textbf{Other Planetary Fates} & & & & \\
        \quad Cold Jupiter (No Excitation)& 36.60\% & 37.95\% & 39.69\% & 35.32\% \\
        \quad Planet Collided on Star & 0.69\% & 1.14\% & 1.36\% & 1.16\% \\
        \hline
\end{tabular}
    \caption{Integrated statistics on the final dynamical fates of planets and their moons following high-eccentricity migration. The systems are hierarchically grouped by the final state of the giant planet: forming a Hot Jupiter, becoming a Free-Floating Planet (Escaped Jupiter), remaining unexcited or torn by the star. The mass threshold separating large and small moons is set to $10^{-5} M_{\mathrm{planet}}$. Percentages may not sum exactly to 100.00\% due to rounding.}
    \label{tab2} 
\end{table*}

For the planets that successfully evolve into hot Jupiters, the most prominent result is the devastation of moon systems during successful high-eccentricity migration: not a single moon regardless of its mass or orbital direction remains stable around a newly formed hot Jupiter ($0.00\%$ survival rate).

To clearly illustrate how this destabilization unfolds, the middle and bottom panels of Fig.~\ref{fig4} detail the satellite dynamics in the representative system introduced in Section~\ref{sec:3.2.1}.
The bottom panel displays the mutual inclination evolution, where the secular perturbations from the outer perturber drive large-amplitude coupled oscillations between the planet's inclination ($i_{\mathrm{p}}$) and the retrograde moon's inclination ($i_{\mathrm{m}}$).
Crucially, the moon's inclination deviates persistently from a purely retrograde co-planar configuration ($i_{\mathrm{m}} \approx 180^{\circ}$), fluctuating down to $\sim 170^{\circ}$.
This inclination departure, coupled with the surge in the host planet's eccentricity toward its peak ($e_{\mathrm{p}} \approx 0.985$, which compresses the pericentre distance), causes the theoretical stability boundary (the reduced Hill sphere, green curve in the middle panel) to collapse drastically.
When this shrinking stability boundary contracts down to the moon's orbital separation, external tidal perturbations rapidly pump the satellite's eccentricity, as evidenced by the broadening envelope in the middle panel.
Consequently, the satellite becomes dynamically unstable before the planet begins its rapid circularization, ultimately resulting in the moon becoming unbound or disrupted (the specific detachment epoch into an unbound orbit is marked by the red cross).

Dynamically, this universal instability is primarily governed by the shrink of the planetary gravitational dominance near pericentre. 
For prograde moons, this complete disruption is a physical inevitability. As indicated by Equation~\ref{eq:A2}, the effective Hill sphere for prograde moons is intrinsically only about half the size of that for retrograde ones. More crucially, when the planetary orbital eccentricity is sufficiently excited to its extreme peak, this reduced Hill sphere shrinks so drastically that it becomes smaller than the physical radius of the host planet itself. Consequently, the stable dynamical region ceases to exist entirely.
Although retrograde moons enjoy a wider theoretical stability region, the violent stellar tidal perturbations experienced during deep pericentre passages persistently excite the moon's eccentricity, ultimately driving retrograde companions either into colliding with the planet's envelope or becoming gravitationally unbound.

\begin{figure*}  
    \centering
    \includegraphics[width=0.95\linewidth]{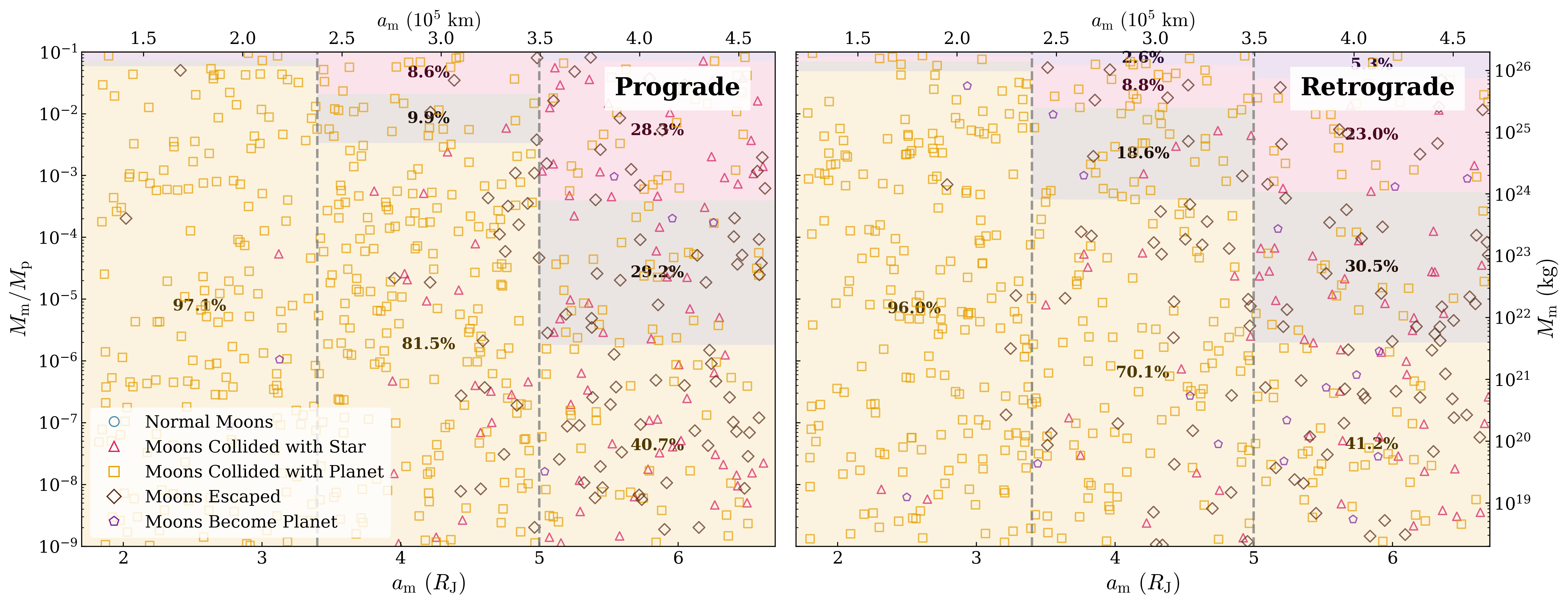}
    \caption{\textbf{Dynamical fates of prograde and retrograde exomoons during planetary high-eccentricity migration.}
    The scatter plot displays the parameter space of the initial semi-major axis ($a_{\mathrm{m}}$) and mass ratio ($M_{\mathrm{m}}/M_{\mathrm{p}}$) for moons orbiting gas giants that eventually become hot Jupiters.
    The left panel illustrates the outcomes for prograde moons, while the right panel shows the retrograde counterparts.
    Distinct markers denote the diverse final fates: green circles for normal stable moons, orange triangles for moons colliding with the central star, magenta squares for moons colliding with the planet, brown diamonds for moons escaping the planetary system, and purple pentagons for moons transitioning into independent planets.
    To statistically quantify these dynamical outcomes, the parameter space is divided into three vertical blocks.
    Within each block, the background is painted with stacked color bands that explicitly represent the fractional distribution of different fates. From bottom to the top, light pink represents moons collided with star, and light yellow for moons collided with planet, light brown for moons escaped, light purple for moons become planets.
    The specific percentages of the dominant outcomes are overlaid directly onto their corresponding color bands to provide a clear quantitative reference.}
    \label{fig5}
\end{figure*}

As detailed in Table~\ref{tab2}, the dominant destruction pathway is direct collision with the host planet, which accounts for nearly all prograde moon fatalities and two-thirds of retrograde moon destruction.
Other minority outcomes for retrograde moons include escaping the stellar system entirely ($ 7.3\%$ to $9.6\%$), colliding with the host star ($4.3\%$ to $7.8\%$), or transitioning into independent heliocentric planets ($ 0.5\%$ to $1.8\%$).
These aggregate statistics are visually corroborated by Figure.~\ref{fig5}, which demonstrates a pronounced orbital distance dependency across different initial semi-major axes ($a_{\mathrm{moon}}$):
\begin{enumerate}
    \item \textbf{Close-in region ($a_{\mathrm{moon}} \lesssim 3.4\,R_{\mathrm{J}}$):} These moons reside deep within the planetary potential well. As the host planet's eccentricity surges, the reduced Hill sphere collapses below the moon's orbital distance, while strong gravitational perturbations rapidly excite the moon's eccentricity. Due to their close initial proximity to the planet, the moons are almost exclusively engulfed by the planetary envelope ($\ge 96.0\%$) before external stellar forces have sufficient time to gravitationally detach or eject them.
    \item \textbf{Intermediate region ($3.4\,R_{\mathrm{J}} < a_{\mathrm{moon}} \le 5.0\,R_{\mathrm{J}}$):} Planetary collision remains the primary outcome ($70.1\%$ to $81.5\%$), but the larger orbital separation allows a modest fraction of moons to be gravitationally stripped near the boundary of the contracting Hill sphere.
    \item \textbf{Outer region ($a_{\mathrm{moon}} > 5.0\,R_{\mathrm{J}}$):} Although distant moons also experience eccentricity excitation, they maintain a relatively large physical separation from the planet's surface while residing close to the outer edge of the Hill sphere. As the Hill sphere contracts, these moons are promptly stripped from the planet's gravitational control. Following this detachment, the liberated moon enters a heliocentric orbit that is nearly co-orbital with the migrating giant planet. During this subsequent chaotic phase, intense secondary gravitational scattering and close encounters with the giant planet dynamically scatter the moon—either imparting strong velocity kicks that eject it from the stellar system entirely ($\sim 29\% - 31\%$), deflecting it onto star-grazing orbits that cause stellar engulfment ($\sim 23\% - 28\%$), or decoupling it into an independent heliocentric planet ($\sim 1\% - 2\%$), while direct planetary collisions drop significantly to $\sim 41\%$. 
\end{enumerate}

\begin{figure}
    \centering
    \includegraphics[width=\linewidth]{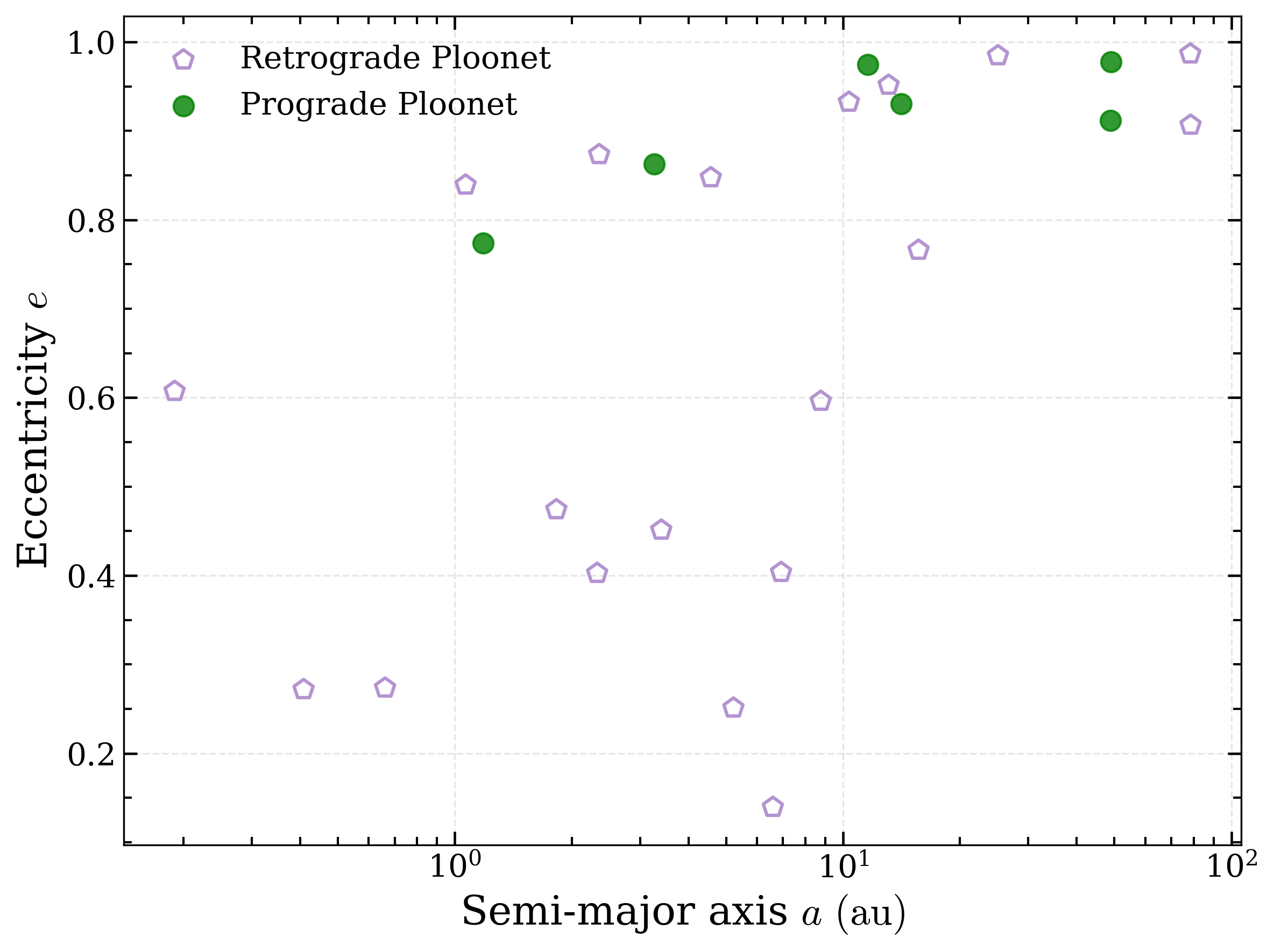}
    \caption{Final Orbits distribution ($a$--$e$) of exomoons escaping from planetary system (ploonets) during coplanar eccentricity excitation. Solid green circles and open purple pentagons denote ploonets originating from prograde and retrograde moon orbits, respectively. These ploonets usually acquire expanded heliocentric orbits than their hosts ($a \gtrsim 1\,\mathrm{AU}$) with moderate-to-high eccentricities.}
    \label{fig6}
\end{figure}

A minor fraction of the exomoons ($\sim 0.5\% - 1.8\%$) decouple from their host planets during the eccentricity excitation phase and transition into independent ploonets.
As illustrated in Figure.~\ref{fig6}, these newly formed ploonets exhibit semi-major axes that are systematically broader than the progenitor planet's initial orbit ($a \gtrsim 1\,\mathrm{AU}$, spanning up to several tens of $\mathrm{AU}$) and possess moderately high orbital eccentricities ($e \sim 0.2 - 0.98$). 
However, their pericentre distances all exceed $0.1 \mathrm{AU}$, making it impossible to initiate high-eccentricity migration or undergo tidal circularization.
Consequently, as their pericentre and apocentre broadly intersect the orbital domain of the eccentric outer perturber, these ploonets are subject to repeated strong gravitational encounters with the outer gas giant. 
We thus infer that the majority of these ploonet configurations represent a transient dynamical state that will inevitably undergo catastrophic multi-body scattering and eventual ejection from the system on $\sim \mathrm{Myr}$ timescales.
Notably, ploonets originating from prograde moons are considerably scarcer than those from retrograde orbits. This asymmetry fundamentally stems from the velocity vector superposition during the detachment phase. Prograde moons possess orbital velocities that additively align with the host planet's heliocentric velocity. Consequently, upon leaving the Hill sphere, they acquire excessive kinetic energy and are far more likely to be entirely ejected from the stellar system or perturbed into star-grazing orbits, whereas the subtractive velocity of retrograde moons allows them to remain marginally bound to the star as ploonets.

Despite the utter destruction of moons around hot Jupiters, Table~\ref{tab2} reveals a counterintuitive phenomenon regarding the ejected free-floating planets (FFPs). 
We find that a remarkably high fraction, roughly $40\%$ of these scattered gas giants successfully retain their moons, carrying them into interstellar space as bound pairs. 
Given this high retention efficiency during violent ejection events, we dedicate a detailed investigation in Section \ref{sec:3.2.3} to uncover the precise dynamical mechanisms that facilitate moon survival around FFPs.

\subsubsection{Free-floating Planets from Coplanar Excitation and Stability of Their Moons}\label{sec:3.2.3}

As noted in Section \ref{sec:3.2.2}, the coplanar high-eccentricity migration mechanism occasionally triggers multi-body scattering, ejecting $10\%$ of the inner gas giants to form free-floating planets.
Remarkably, our simulations reveal that these violent ejections do not inherently strip away all the exomoons.
Between $35.6\%$ and $49.2\%$ of these FFPs successfully retain their moons, meaning that $3.6\%$ to $5.3\%$ of the total simulated systems end up as bound planet-moon pairs drifting in interstellar space.

\begin{figure*}  
    \centering
    \includegraphics[width=0.95\linewidth]{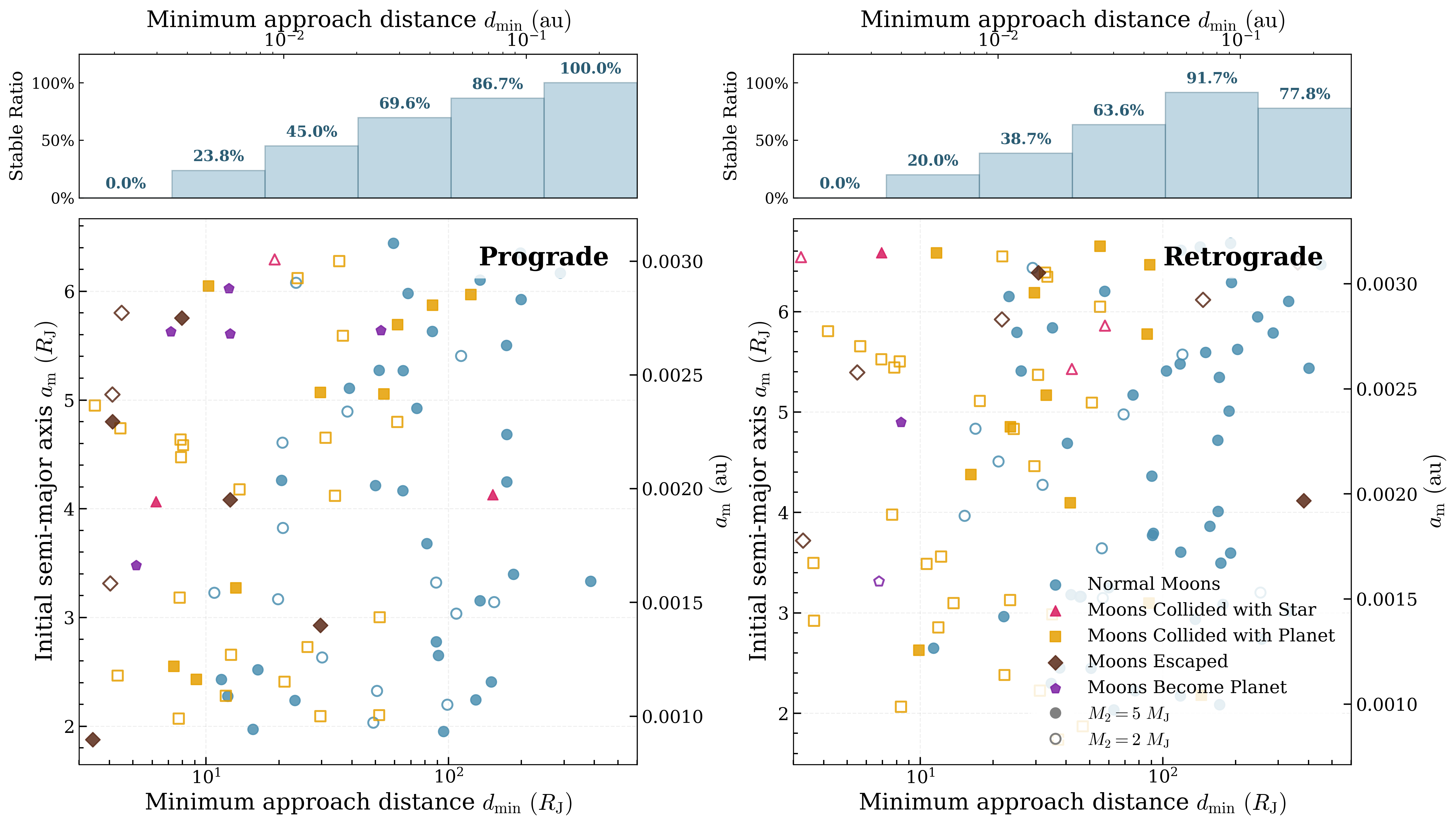}
    \caption{Dynamical fates of exomoons around ejected free-floating planets (FFPs) as a function of their initial semi-major axis ($a_m$) and the minimum approach distance ($d_{\mathrm{min}}$) between the host planet and the outer perturber during the scattering phase. The left and right panels display the outcomes for prograde and retrograde moons, respectively. Distinct markers and colors denote different final outcomes: stable moons successfully retained by the FFP (blue circles), collisions with the host planet (yellow squares), collisions with the central star (pink triangles), ejections from the stellar system (brown diamonds), and decoupling into independent ploonets (purple pentagons). Solid and open symbols differentiate the mass of the outer perturbing cold Jupiter, corresponding to $M_{\mathrm{2}} = 5\,M_{\mathrm{J}}$ and $2\,M_{\mathrm{J}}$, respectively. Stable planet--moon pairs systematically cluster in regions with larger $d_{\mathrm{min}}$, indicating that moon survival during planetary ejection strongly depends on avoiding deep, penetrating encounters.}
    \label{fig7}
\end{figure*}

To elucidate the dynamical conditions that permit moon survival during close encounters, we trace the minimum approach distance ($d_{\mathrm{min}}$) between the inner planet and the massive outer planet throughout the entire evolution of the system.
Figure.~\ref{fig7} illustrates the dynamical fates of exomoons within the phase space of this global $d_{\mathrm{min}}$ and their initial semi-major axes ($a_m$).
A distinct physical correlation emerges between the historical closeness of the planetary encounters and the ultimate fate of the moon. 
Stable moons (blue circles) systematically cluster in regions with larger global minimum approach distances ($d_{\mathrm{min}} \gtrsim 30\,R_{\mathrm{J}}$). 
This indicates that the host planets in these surviving systems avoided ultra-close, penetrating encounters throughout their evolution. 
Because the outer perturber's tidal field is strongly distance-dependent, a universally large $d_{\mathrm{min}}$ ensures that the maximum perturbative shear never breaches the planet's Hill sphere. Consequently, moons can survive the entire ejection process intact---even those on relatively wide initial orbits ($a_m$ up to $\sim 7\,R_{\mathrm{J}}$), provided $d_{\mathrm{min}}$ remains sufficiently large.

Conversely, systems that experience close planetary encounters (smaller $d_{\mathrm{min}}$) subject their moons to intense gravitational perturbations. 
In this regime, the perturbation is strong enough to deeply penetrate the circumplanetary environment, meaning that the moon's initial semi-major axis ($a_m$) no longer guarantees its survival. 
As shown in Figure.~\ref{fig7}, destabilized moons across the entire simulated spatial range suffer a variety of outcomes: colliding with the host planet (yellow squares) or the central star (pink triangles), escaping the system (brown diamonds), or becoming independent ploonets (purple pentagons). 
Because planet-planet scattering is a rapid dynamical process, the specific final fate of a destabilized moon is highly stochastic, likely depending on its exact orbital phase at the moment of closest approach.

Additionally, the mass of the outer perturber directly dictates the spatial scale of the planetary ejections. 
Because a more massive companion exerts a significantly stronger gravitational perturbation, it is capable of scattering the inner gas giant out of the system from a much greater distance. 
As demonstrated by the solid symbols in Figure.~\ref{fig7}, ejections driven by a $5\,M_{\mathrm{J}}$ perturber can readily occur even when the minimum approach distance is as large as $d_{\mathrm{min}} \sim 200\,R_{\mathrm{J}}$.
In contrast, when the perturber is less massive ($2\,M_{\mathrm{J}}$, open symbols), its reduced gravitational reach necessitates a much deeper penetration into the inner system, with the scattering events predominantly occurring at closer distances of $d_{\mathrm{min}} \lesssim 30\,R_{\mathrm{J}}$.

\vspace{-1em}
\section{Discussion}

\subsection{Strong Moon Shielding Effect in Coplanar Excitation}\label{sec:4.1}

\citet{2020MNRAS.499.4195T} demonstrated that a massive exomoon ($\sim0.1M_{J}$) could dynamically shield its host planet from external perturbations.
In their analytical framework, this shielding condition was formulated by comparing
the Kozai-Lidov timescale of the inner moon-planet system $T_{\mathrm{quad,inn}}$ with the quadrupole timescale of the stellar perturber $T_{\mathrm{quad,out}}$ (e.g. their Eq.~13).
However, while they qualitatively noted that the shielding originates from the apsidal precession induced by the moon's quadrupole moment, evaluating it via $T_{\mathrm{quad,inn}}$ yields a scaling that moon-shielding effect is slightly negative correlation to the moon's mass, which appears inconsistent with the expectation that more massive moons should exert stronger dynamical influence.

Here, we suggest a refined analytical formulation based on the standard secular perturbation theory of hierarchical triples \citep{2000ApJ...535..385F,2015MNRAS.449.4221H}.
The shielding effect is fundamentally governed by the apsidal precession timescale of the planetary orbit $T_{\mathrm{prec}}$ induced by the moon.
For $M_{\mathrm{moon}} \ll M_{\mathrm{planet}}$, this precession timescale scales strictly as $T_{\mathrm{prec}} \propto 1/M_{\mathrm{moon}}$.
In coplanar configurations, the high-eccentricity excitation differs from the highly inclined ZLK mechanism.
As noted by \citet{2015MNRAS.452.3610A} and \citet{2020MNRAS.499.4195T}, coplanar configurations exhibit no angular momentum exchange at the quadrupole level.
Instead, the growth of planetary eccentricity is exclusively driven by slower octupole-order perturbations.
Therefore, the coplanar shielding condition should be evaluated by comparing $T_{\mathrm{prec}}$ with the perturber's outer octupole timescale $T_{\mathrm{oct,out}}$, where $T_{\mathrm{oct,out}} = T_{\mathrm{quad,out}} / \sqrt{\epsilon_{\mathrm{oct}}}$.
We thus define the dimensionless coplanar shielding parameter $\mathcal{R}_{\mathrm{coplanar}}$ as:
{\allowdisplaybreaks
\begin{align}
T_{\mathrm{prec}} &\approx \frac{4}{3} P_{\mathrm{p}} \left(\frac{M_{\mathrm{p}}}{M_{\mathrm{m}}}\right) \left(\frac{a_{\mathrm{p}}}{a_{\mathrm{m}}}\right)^2 (1 - e_{\mathrm{1}}^2)^2 \nonumber \\
T_{\mathrm{quad,out}} &\approx P_{\mathrm{p}} \left(\frac{M_\star}{M_{\mathrm{2}}}\right) \left(\frac{a_{\mathrm{2}}}{a_{\mathrm{p}}}\right)^3 (1 - e_{\mathrm{2}}^2)^{3/2} \nonumber \\
\epsilon_{\mathrm{oct}} &= \frac{a_{\mathrm{p}}}{a_{\mathrm{2}}} \frac{e_{\mathrm{2}}}{1 - e_{\mathrm{2}}^2} \nonumber \\
\mathcal{R}_{\mathrm{coplanar}} &= \frac{T_{\mathrm{prec}}}{T_{\mathrm{oct,out}}} = \frac{T_{\mathrm{prec}}}{T_{\mathrm{quad,out}}} \sqrt{\epsilon_{\mathrm{oct}}}.
\end{align}}
This revised framework naturally resolves the mass-dependence discrepancy, demonstrating that a more massive moon drives faster apsidal precession, thereby enhancing its shielding capability.
Furthermore, because the octupole timescale is inherently longer than the quadrupole one ($\sqrt{\epsilon_{\mathrm{oct}}} < 1$), this formulation confirms that moons can provide significantly stronger shielding against coplanar planetary perturbers than against highly inclined stellar ones.

To quantify whether this theoretical shielding translates effectively to our specific parameter space, we performed another set of simulations comprising three groups of 200 systems each.
In this setup, the external perturber is fixed as a $5\,M_{\mathrm{J}}$ planet on a eccentric orbit ($a_{2} = 6\,\mathrm{AU}$, $e_{2} = 0.7$).
For the moon configurations, we placed the moons on relatively wide retrograde orbits at $a_{m} = 360,000\,\mathrm{km}$ ($\sim 5\,R_{\mathrm{J}}$).
The three groups correspond to a control group (no moon), a small moon group ($M_{moon} = 2 \times 10^{18}\,\mathrm{kg}$), and a massive moon group ($M_{moon} = 6 \times 10^{25}\,\mathrm{kg}$, roughly $10\,M_{\oplus}$).
Each system was integrated for $10^5$ years, unless the host planet successfully migrated into a hot Jupiter, collided with the central star, or was dynamically ejected.
We then tracked the absolute cumulative fraction of hot Jupiters across these three scenarios.

\begin{figure}
    \centering
    \includegraphics[width=\linewidth]{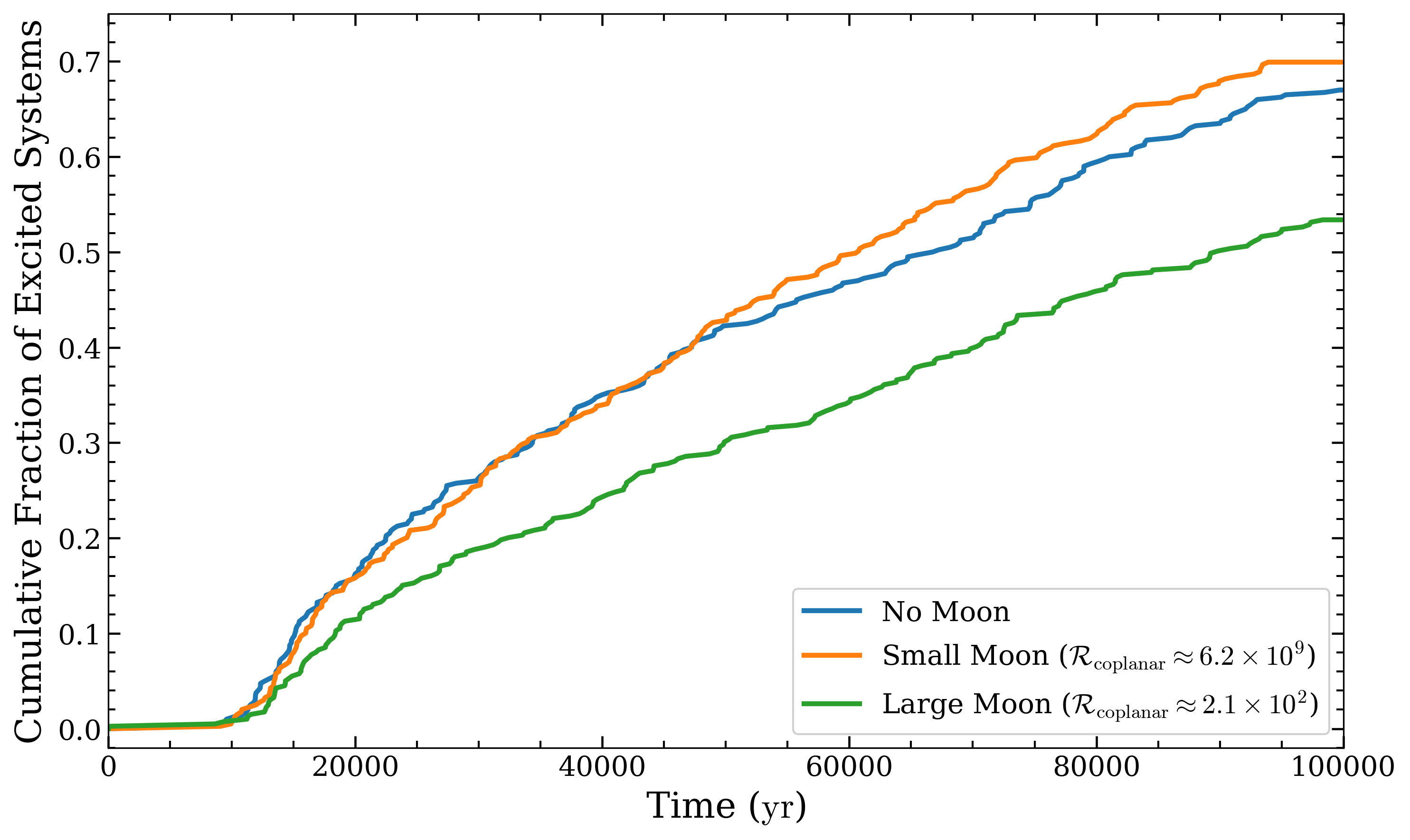}
    \caption{The absolute cumulative fraction of planetary eccentricity excitation over time for different moon configurations. The blue line represents the control group with no moons. The orange line corresponds to systems hosting a minor moon ($m = 2 \times 10^{18}\,\mathrm{kg}$), which shows negligible deviation from the moonless scenario. The green line denotes systems hosting a massive moon ($m = 6 \times 10^{25}\,\mathrm{kg}$, corresponding to $10\,M_{\oplus}$), which effectively acts as a dynamical shield and reduces the overall probability of high-eccentricity migration by approximately $30\%$.}
    \label{fig8}
\end{figure}

In Fig.~\ref{fig8}, the shielding effect exhibits a strict mass threshold that aligns closely with the theoretical timescale criteria.
First, the dynamical trajectory of the small-moon configuration ($m = 3 \times 10^{18}\,\mathrm{kg}$, orange curve) is virtually indistinguishable from that of the bare planetary control group (blue curve). 
Both curves ascend at nearly identical rates and converge to final excitation fractions of approximately $70\%$. 
This confirms that small moons possess inner quadrupole timescales far too long to satisfy the critical relation, thereby offering zero protection against external octupole-driven resonance.
In stark contrast, systems hosting a massive moon ($m = 6 \times 10^{25}\,\mathrm{kg}$, green curve) deviate significantly from the control group, demonstrating clear statistical suppression of high-eccentricity migration. 
While the baseline systems achieve an ultimate excitation probability of around $70\%$, the massive-moon cohort is capped at roughly $50\%$. 
This translates to a robust absolute reduction of approximately $30\%$ in the overall yield of hot Jupiters. 
This demonstrates that the gravitational and tidal contributions from a massive moon can alter the secular dynamics of coplanar configurations and hinder extreme orbital evolution.

In this formulation, $\mathcal{R}_{\mathrm{coplanar}} \ll 1$ represents the stringent adiabatic limit where external octupole perturbations are completely suppressed. 
However, for our massive moon scenario, evaluating Eq.~(7) yields $\mathcal{R}_{\mathrm{coplanar}} \approx 210$ (compared to $\approx 6.2 \times 10^9$ for the minor moon). 
Despite $\mathcal{R}_{\mathrm{coplanar}} > 1$, the massive moon successfully provides noticeable dynamical shielding. 
This occurs because high-eccentricity excitation relies on secular resonant libration; even a moderately shortened $T_{\mathrm{prec}}$ introduces an independent precession frequency that detunes this delicate resonance, driving the argument of pericentre into circulation and preventing extreme eccentricity growth. 
Consequently, an exomoon can substantially disrupt secular excitation well before reaching the formal adiabatic threshold $\mathcal{R}_{\mathrm{coplanar}} \ll 1$.

\subsection{Comparison with Hot Jupiters produced by Other mechanisms}\label{sec:4.2}

We have shown that coplanar high-eccentricity migration (CHEM) serves as an efficient pathway to produce low-inclination hot Jupiters. 
In our simulations, the planetary eccentricity is secularly excited to extreme values (e.g., $e \gtrsim 0.98$), culminating in a rapid tidal circularization.
Interestingly, our resulting final semi-major axes ($a_f \lesssim 0.035 \mathrm{\ AU}$) are systematically tighter than those predicted by \citet[e.g., $a_f \sim 0.05 \mathrm{\ AU}$ in][]{2015ApJ...805...75P,2026arXiv260620803G}. 
We emphasize that this discrepancy originates fundamentally from the choice of tidal dissipation models. 
Our simulations employ a dynamical tide model, within which a larger final semi-major axis could only be achieved if the planet possessed a significantly larger inflated radius ($R_{\mathrm{p}}$) or a higher dynamical tide efficiency factor ($f_{\text{dyn}}$). 
Conversely, the equilibrium tide model utilized in previous studies relies heavily on the constant time-lag parameter ($\tau$). 
Adjusting this viscous time-lag can easily yield a wider circularization radius consistent with their results. 
Therefore, placing tighter observational and theoretical constraints on tidal models is crucial for accurately determining the final parking locations of hot Jupiters.

Crucially, our simulations demonstrate that no moon can survive the coplanar high-eccentricity migration once the extreme eccentricity is reached.
Instead, massive exomoons can act as a dynamical shield to halt the eccentricity excitation.
To place our findings in a broader dynamical context, we synthesize and contrast the moon survival rates and shielding efficiency across major hot Jupiter formation channels in Table~\ref{tab3}, and discuss their respective physical regimes below.

In the highly inclined von ZLK mechanism, \citet{2020MNRAS.499.4195T} has demonstrated primordial exomoons are systematically destroyed during the excitation phase, and $10\%$ of Neptune-sized exomoons could shield the planet from being perturbed.
Similarly, other mechanisms, such as the E1 and E2 scenarios proposed by \citet{2017ApJ...848...20W}, are expected to be equally destructive.
In these pathways, planetary orbits experience eccentricity modulation coupled with Kozai-like inclination growth (E1) and sudden prograde-retrograde flipping (E2).
The intense multi-body interactions and violent orbital variations in these configurations leave no stable space for moons, likely overwhelming any dynamical shielding effect from massive moons.
Conversely, for multi-planet systems governed by secular chaos \citep{1989Natur.338..237L,1997A&A...317L..75L}, the growth of eccentricity is driven by the overlap of higher-order resonances over much longer timescales \citep{2011ApJ...735..109W}. 
Theoretically predicting exomoon survival or moon shielding efficacy in this regime is highly challenging because the multi-frequency phase-space diffusion makes the long-term dynamics fundamentally unpredictable, and the exceedingly slow excitation efficiency hampers direct $N$-body simulations of circumplanetary orbits.

Conversely, planet-planet scattering (PPS) operates through impulsive close encounters \citep{2008ApJ...678..498N, 2012ApJ...751..119B}. 
During the initial violent scattering phase, exomoons are highly susceptible to being stripped; for instance, \citet{2018ApJ...852...85H} demonstrated that moons orbiting beyond $0.1\,R_{\mathrm{H}}$ are systematically destroyed. 
In addition to reshaping the bound planetary architecture, PPS is highly efficient at generating free-floating planets (FFPs). 
Recent studies, such as \citet{2026ApJ...998..245H}, systematically evaluated exomoon survival during these PPS-induced ejections, concluding that survival is strictly dictated by the minimum planet-planet approach distance ($d_{\mathrm{min}}$). 
Importantly, systems with more massive perturbing planets can scatter planets at larger $d_{\mathrm{min}}$, hence experience fewer deep encounters, resulting significantly higher moon survival rates.

Our findings are consistent with this physical picture.
Specifically, for Jupiter-like systems, previous scattering experiments have established a critical safety threshold: encounters occurring beyond roughly $0.03 \mathrm{\ AU}$ exert negligible dynamical perturbations on compact Galilean-like moon systems \citep{2014AJ....148...25D}. 
As illustrated in Fig.~\ref{fig7}, our simulations clearly reproduce this $0.03 \mathrm{\ AU}$ survival boundary. 
Although FFPs represent a byproduct ($\sim 10\%$) in our coplanar scatterings, roughly $40\%$ of these ejected giants successfully retain their massive moons.
Ejections driven by a more massive perturber ($5\,M_{\mathrm{J}}$) typically occur at larger distances ($d_{\mathrm{min}} \sim 0.06 \mathrm{\ AU}$), safely preserving the circumplanetary orbits. 
In contrast, ejections by a less massive perturber ($2\,M_{\mathrm{J}}$) require deeply penetrating close encounters ($d_{\mathrm{min}} < 0.03 \mathrm{\ AU}$) that systematically strip or destroy the moons. 
Ultimately, a larger perturber mass facilitates more distant, gentle ejections, allowing the planet to safely carry its moon into interstellar space.

\begin{table*}
\centering
\small
\setlength{\tabcolsep}{4pt}
\caption{Comparison of Exomoon Fate and Moon-Shielding Feasibility Across Hot Jupiter Formation Channels}
\label{tab3}
\begin{tabularx}{\textwidth}{>{\raggedright\arraybackslash}p{3.2cm} >{\raggedright\arraybackslash}X >{\centering\arraybackslash}p{2.3cm} >{\raggedright\arraybackslash}p{2.8cm} >{\raggedright\arraybackslash}X}
\hline\hline
HJ Formation Channel & Perturber Type \& Configuration & Moon Survival Rate & Moon Shielding Efficacy & Key Literature Reference \\
\hline
\textbf{Type-II Disk Migration} & N/A & $\sim 8\%$ (Pro) / $\sim 24\%$ (Retro) & N/A & This work; \citet{2010ApJ...719L.145N}\\
\textbf{Coplanar High-$e$ (CHEM)} & Coplanar eccentric giant & 0.00\% & $\sim 30\%$ ($ 10\,M_{\oplus}$ moon) & This work; \citet{2015ApJ...805...75P} \\
\textbf{von Zeipel--Lidov--Kozai} & Inclined stellar binary & 0.00\% & $\sim 10\%$ ($ 0.1\,M_{\mathrm{J}}$ moon) & \citet{2020MNRAS.499.4195T} \\
\textbf{Planet--Planet Scattering} & Multiple giant planets (Impulsive) & Stripped ($>0.1\,R_{\mathrm{H}}$) & Unlikely & \citet{2018ApJ...852...85H}; \citet{2026ApJ...998..245H} \\
\textbf{Inclination Flipping} & Coplanar / inclined multi-planets & Predicted 0\% & Unlikely & \citet{2017ApJ...848...20W} \\
\textbf{Secular Chaos} & Resonant multi-planet diffusion & Unconstrained & Unconstrained & \citet{2011ApJ...735..109W} \\
\hline\hline
\end{tabularx}
\begin{flushleft}
\footnotesize
\textit{Note:} Survival rates reflect primordial retention through orbital circularization before long-term tidal/magnetic decay. For inclination flipping (E1/E2) and secular chaos, exact survival rates have not been modeled with circumplanetary test particles and represent theoretical predictions.
\end{flushleft}
\end{table*}

\subsection{Other Astrophysical Barriers to Observe Exomoons Around Hot Jupiters}\label{sec:4.3}

Our N-body simulations reveal a clear dependence of moon survival on the planetary migration mechanism. 
While high-eccentricity migration completely strips all primordial moons (Sec.~\ref{sec:3.2.2}), disk migration allows close-in companions to survive, specifically prograde moons within $\sim 3\,R_{\mathrm{J}}$ and retrograde moons within $\sim 6\,R_{\mathrm{J}}$ (Sec.~\ref{sec:3.1}). 
However, this dynamical survival does not ensure that such exomoons can be observationally confirmed around hot Jupiters.
Current transit photometry and transit timing variation (TTV) methods favor the detection of massive moons, typically requiring an Earth- to Neptune-mass moon for a robust signal \citep{2013ApJ...777..134K, 2018SciA....4.1784T}. 
Even if a moon successfully survives disk migration, forming a sufficiently massive moon around the gas giant and retaining it on Myr timescales are hindered by three additional astrophysical barriers.

Firstly, forming an adequately massive moon around a gas giant is intrinsically difficult, primarily because native formation mechanisms, namely
in-situ formation \citep[e.g.][]{2003Icar..163..198M,2006Natur.441..834C,2012Sci...338.1052C}, and giant impacts \citep[e.g.][]{1975Icar...24..504H} are highly inefficient in these environments.
In-situ models indicate that Jupiter-mass planets at $1 ~\mathrm{AU}$ or $5 ~\mathrm{AU}$ 
cannot easily assemble moons larger than Mars \citep[e.g.][]{2012ApJ...753...60O,2020MNRAS.492.5336R}.
Planetary illumination and viscous heating deplete the local ice inventory, severely reducing the available solid mass in the circumplanetary disk.
Furthermore, any massive moons that do manage to accrete tend to form outside the snow line, far from the planet \citep{2012ApJ...753...60O,2015A&A...578A..19H}, leaving them highly vulnerable to being stripped when the Hill sphere shrinks during inward migration.
Giant impacts face similar obstacles.
Circumplanetary disks around gas giants incorporate thick hydrogen and helium envelopes, and impacts involving large rocky planets ($>6\,M_{E}$) or icy ($>1\,M_{E}$) progenitors generate disks rich in vapor. This vapor exerts strong aerodynamic drag on moonlets, forcing them to rapidly spiral into the host planet before accumulating substantial mass \citep{2020MNRAS.492.5336R,2022NatCo..13..568N}.

Alternatively, a planet could acquire a massive companion through gravitational capture, analogous to Triton's origin \citep{2006Natur.441..192A}.
Indeed, \citet{2018A&A...610A..39H} claimed that a Neptune-mass object could be captured by a super Jupiter like Kepler-1625 b.
However, such capture events are dynamically suppressed prior to planetary migration.
The surrounding protoplanetary disk efficiently circularizes planetary orbits and heavily inhibits the close encounters required for a successful capture.
Consequently, forming a massive moon around a hot Jupiter progenitor is highly improbable, and any viable capture scenario would inherently occur late in the system's evolution, typically after the protoplanetary disk has dissipated and migration has concluded.

Secondly, even if a massive moon successfully migrates with its host planet, extreme tidal heating and stellar radiation will drastically erode its observability. 
Close-in moons around hot Jupiters experience intense internal heating and atmospheric evaporation. \citet{2009ApJ...704.1341C} demonstrated that moons with volatile-rich or thick envelopes undergo severe atmospheric stripping if the host planet's orbital period is less than 6.6 days.
This profound erosion efficiently reduces the physical radius of super-Earths or sub-Neptunes, plummeting their transit depths well below current instrumental detection thresholds.

Finally, non-gravitational dynamical forces act as an inescapable terminal barrier. 
As demonstrated by our disk migration results, surviving exomoons are confined to compact orbits of $\sim 3-6\,R_{\mathrm{J}}$, corresponding to roughly $0.04\,R_{\mathrm{h}}$ for a mature hot Jupiter. 
At this extreme proximity, the moon's orbit resides strictly within the planetary corotation radius, where strong planet-moon magnetic interactions dominate the long-term orbital evolution \citep{2017ApJ...847L..16S}. 
While icy and silicate moons might initially resist these magnetic torques, icy moons accreted beyond the snow line will inevitably lose their volatile water content during the planet's inward migration, elevating their electrical conductivity and the magnetic drag. 
Consequently, the intense magnetic torque rapidly extracts angular momentum from the moon, causing it to spiral inward and cross the planetary Roche limit on a remarkably short timescale of just $\sim 3\,\mathrm{Myr}$ \citep{2024ApJ...965...88W}. 
Therefore, even if a massive close-in moon survives the initial dynamical violence, it will only remain transiently stable immediately after the hot Jupiter completes its migration before facing inevitable physical destruction.

In summary, while pure N-body dynamics allow for limited moon survival during disk migration, the actual observation of such exomoons is prevented by a sequence of physical barriers. 
First, assembling an observationally massive moon is inherently difficult due to hostile circumplanetary disk conditions, while dynamical capture is strongly suppressed prior to migration. 
Second, any massive moon that successfully co-migrates will suffer severe atmospheric stripping from intense stellar radiation and tidal heating, effectively reducing its radius below current detection thresholds. 
Finally, the surviving compact orbits inevitably place these moons within the planetary corotation radius, where magnetic torques will drive them into the Roche limit and destroy them within a few million years. 
Together, these compounding effects ensure that the prospects for detecting primordial moons around mature hot Jupiters are exceedingly dim.
In contrast to the hostile, barrier-ridden environments of close-in hot Jupiters, cold Jupiters and free-floating planets (FFPs) are largely immune to such stellar proximity hazards.
Consequently, cold Jupiters observed via transit and timing techniques, alongside FFPs detectable via planetary gravitational microlensing events \citep[e.g.,][]{2010A&A...520A..68L}, offer the most viable and promising observational avenues for future exomoon searches.

\section{Conclusion}

The detection and characterization of exomoons provide crucial insights into the formation history and dynamical evolution of planetary systems.
To date, confirmed detections remain elusive, and none of the promising candidates orbit hot Jupiters.
Previous studies primarily investigated the stability of moons around mature hot Jupiters, focusing on long-term tidal decay, magnetic interactions and heating after the migration process was completed.
In contrast, this study concentrates on the survival, stability and ultimate dynamical fates of exomoon during the planetary migration processes themselves, specifically disk migration and coplanar high-eccentricity migration.
Our primary findings are summarized as follows:

\begin{enumerate}

\item Bound exomoon retention around hot Jupiters is confined only in disk migration model. Prograde moons survive only within a narrow stability zone ($\sim 3\,R_{\mathrm{J}}$, $\sim 8\%$ fraction), whereas retrograde moons exhibit a broader safe belt ($\sim 6\,R_{\mathrm{J}}$, $\sim 25\%$ fraction) that remains robust against mass variations and mutual companion scattering. Conversely, in successful CHEM channels, circumplanetary survival drops strictly to 0.00\%, as extreme pericentre Hill-sphere shrinkage and impulsive stellar perturbations completely truncate the circumplanetary stability domain.

\item Unstable satellites follow divergent destruction channels dictated primarily by their orbital proximity to the host planet. Close-in moons are predominantly engulfed by the planetary envelope or tidally disrupted as the Hill sphere contracts. In contrast, distant companions are preferentially stripped into heliocentric orbits, transitioning into transient ploonets or escaping the system entirely via secondary multi-body scattering.

\item Massive exomoons ($\sim 10\,M_{\oplus}$) exert profound dynamical feedback on their host planet via rapid inner apsidal precession. In $\sim 30\%$ of CHEM cases, this precession successfully suppresses octupole-order secular perturbation, halting eccentricity excitation and preserving the planet-moon system in its primordial cold-Jupiter orbit.

\item During chaotic multi-body scattering in CHEM, $\sim 10\%$ of inner giants are dynamically ejected into interstellar space as free-floating planets (FFPs), with roughly $40\%$ of them successfully retaining their moons, particularly when encounters avoid penetrating close approaches ($d_{\min} \gtrsim 30\,R_{\mathrm{J}}$).
\end{enumerate}

Combining these dynamical instabilities and physical barriers, we conclude that hot Jupiters are highly unlikely to host observable moons.
Consequently, rather than targeting hot Jupiters, present and future observational missions aiming to detect exomoons including Roman and CSST should focus on the substantial population of cold Jupiters and free-floating planets \citep{2025AJ....170..258L,2025RAA....25f5014F}, which provide a much more stable environment for moon retention.

\section{Acknowledgments}
This work is supported by the National Key R\&D Program of China (No. 2024YFA1611801), Civil Aerospace Technology Research Project (D010102), and and science research grants from the China Manned Space Project (No. CMSCSST-2025-A16). We also appreciate the suggestions of Jiwei Xie at Nanjing University and Shangfei Liu at Sun Yat-sen University. We also thank Keting Xin and Qingfeng Pang at Nanjing University for their assistance with using \textsc{rebound}. This research made use of \textsc{rebound} \citep{2012A&A...537A.128R}, Astropy \citep{2013A&A...558A..33A}, as well as NumPy and SciPy \citep{2011CSE....13b..22V}, and Matplotlib \citep{2007CSE.....9...90H}.



\bibliographystyle{mnras}
\bibliography{reference.bib} 



\appendix

\section{Stability Hill sphere boundary simulation}\label{appendix A}
\hspace*{2em}

The Hill sphere, often referred to as the Roche sphere, defines the roughly spherical region encompassing a celestial body where its gravitational dominance overrides that of the host star.
For a planet, this sphere dictates the spatial limits within which a moon can maintain a dynamically stable orbit. The classical Hill radius, $R_{H}$, is typically expressed as:
\begin{equation}
R_{H}=a_{p}\left(1- e_{p}\right)\left(\frac{M_{p}}{3 M_{*}}\right)^{1 / 3}
\label{KD}
\end{equation}

Previous orbital dynamics studies have demonstrated that moons near the outer edge of the Hill sphere are highly susceptible to long-term instabilities, frequently resulting in collisions with the Roche lobe of the planet or ejection from the system \citep{1967MNRAS.136..245H}.
To quantify this stable fraction, \citet{1967MNRAS.136..245H} established $A = 0.53$ for prograde moons and $A = 0.69$ for retrograde ones. 
Conversely, Hinkel adopted a more conservative $A = 0.3$ for prograde configurations based on the analytical framework of Barnes and O'Brien \citep{2002ApJ...575.1087B,2013ApJ...774...27H}.
Furthermore, \citet{2006MNRAS.373.1227D} derived $A = 0.49$ and $A = 0.93$ for prograde and retrograde orbits, respectively, via rigorous N-body simulations.
These latter values have been extensively utilized to constrain the stability limits of exomoons around known exoplanets, such as HD 100777 b and WASP-49 A b \citep{2022PASJ...74..815G,2025A&A...694L...8S}.

However, these canonical stability boundaries were predominantly derived from numerical simulations tailored to cold Jupiters on distant orbits.
Consequently, these established constants may not accurately capture the extreme tidal and perturbative environments characteristic of hot Jupiters.
Furthermore, existing prescriptions often treat orbital inclination as a discrete variable. A continuous, inclination-dependent formulation for the reduced Hill radius is therefore required to accurately assess moon survivability in extreme close-in systems.

To address this gap, we conduct dedicated numerical simulations to derive an empirical formula for the stable Hill sphere specifically tailored to the hot Jupiter regime.
Evaluating the stability of this restricted three-body problem across a broad parameter space necessitates a robust quantitative diagnostic.
For this purpose, we employ the Mean Exponential Growth factor of Nearby Orbits (MEGNO), a widely utilized chaos indicator in celestial mechanics and non-linear dynamics \citep{2000A&AS..147..205C,2001A&A...378..569G}. 
MEGNO evaluates the degree of chaos by time-averaging the exponential divergence of adjacent phase-space trajectories. 
A MEGNO value converging to $2$ indicates a quasi-periodic, dynamically stable system, whereas a value exceeding $3$ signifies highly chaotic behavior characterized by exponential orbital divergence.

\begin{figure}
    \centering
    \includegraphics[width=\linewidth]{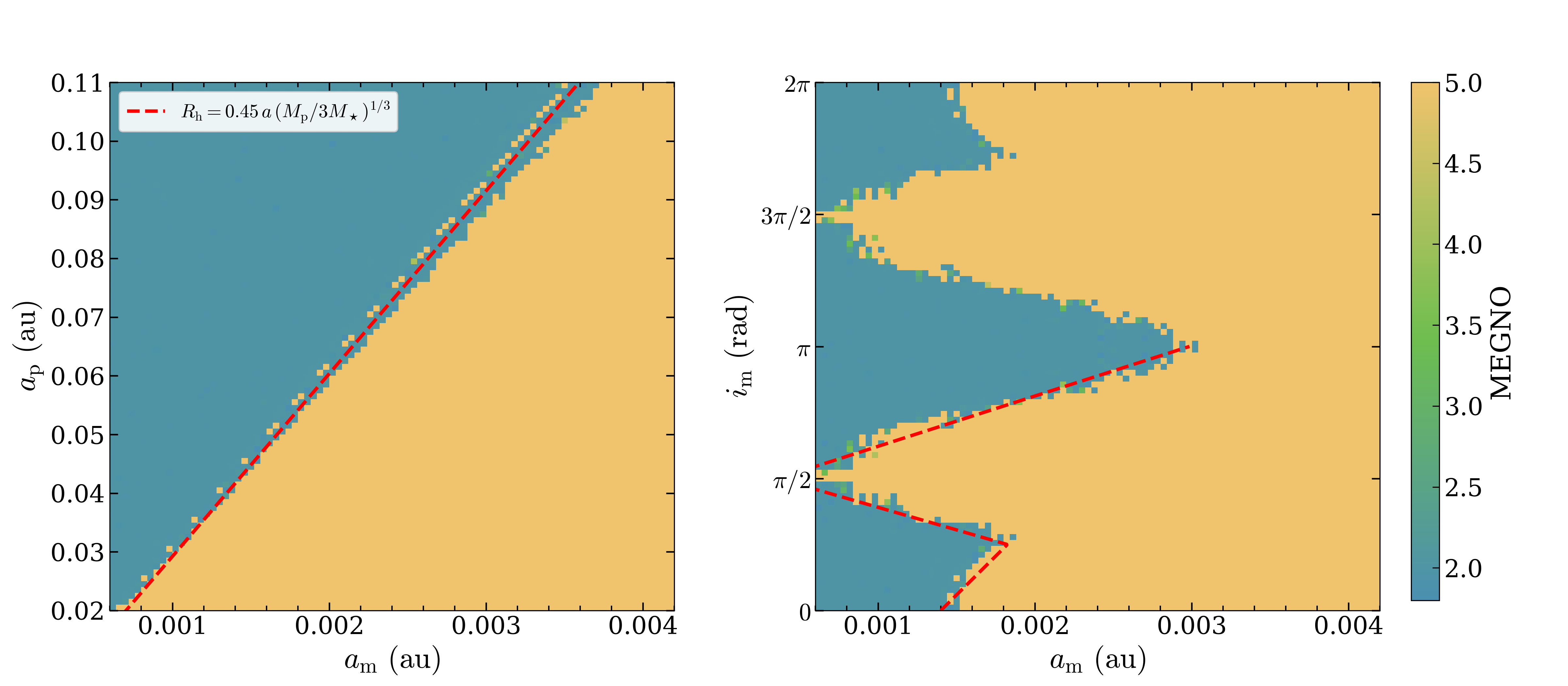}
    \caption{Left: The MEGNO stability map as a function of the moon's initial semi-major axis and the host planet's orbital eccentricity. High MEGNO values (yellow) denote chaotic, unstable regimes, while low values (blue) indicate regular, stable orbits. Right: The MEGNO map illustrating the dependency of moon's stability on the initial semi-major axis and the moon's orbital inclination. Note that the inclination is expressed in radians.}
    \label{fig:two_images}
\end{figure}

Our parameter space exploration primarily focuses on the interplay between the host planet's semi-major axis ($a$),  and the moon's mutual orbital inclination ($i$). 
We first isolate the effects of the planetary orbit by fixing $i=0^{\circ}$ and varying $a$ to delineate the baseline range of the reduced Hill radius. 
Subsequently, we fix the planetary orbital parameters to systematically sweep the inclination $i$. 
Each unique configuration is integrated for 10,000 orbits. 
The resultant stability boundaries from the MEGNO maps are then linearly fitted. By synthesizing these piecewise fits, we derive a comprehensive stability boundary that encapsulates all simulated parameters. The corresponding MEGNO heatmaps and the newly fitted piecewise functions are presented in Figure.~\ref{fig:two_images} and Equation~(\ref{eq:A2}), respectively.

\begin{equation}\label{eq:A2}
\begin{aligned}
R_{h} &= a(1-e)\left(\frac{M_{p}}{3 M_{*}}\right)^{\frac{1}{3}} \cdot I \\
I &= 0.153i + 0.450, \quad i \in \left[0, \frac{\pi}{4}\right] \\
   &= -0.509i + 0.970, \quad i \in \left(\frac{\pi}{4}, \frac{\pi}{2}\right] \\
   &= 0.484i - 0.590, \quad i \in \left(\frac{\pi}{2}, \pi\right]
\end{aligned}
\end{equation}

This continuous, inclination-dependent model serves as our primary criterion for assessing moon's stability throughout the high-eccentricity migration simulations in this work. 
We note that the presence of additional planetary companions could theoretically further truncate the size of the reduced Hill sphere (e.g., yielding $I=0.4$ for prograde moons and $I=0.65$ for retrograde moons) \citep{2013ApJ...775L..44P}. 
However, this secondary perturbation does not compromise our statistical outcomes, as we have adopted a conservatively broad spatial threshold for defining moon's escape in our numerical tracking, which thoroughly encompasses any boundary shrinkage induced by multi-planet interactions.


\bsp	
\label{lastpage}
\end{document}